\documentclass[12pt, a4paper]{article}

\usepackage[english]{babel}
\usepackage[utf8x]{inputenc}
\usepackage[T1]{fontenc}
\usepackage[margin=1in]{geometry}

\usepackage{amsmath, amssymb, amsthm}
\usepackage{graphicx}
\usepackage{setspace}
\usepackage{natbib}
\usepackage{bm, bbm}
\usepackage[colorinlistoftodos]{todonotes}
\usepackage[colorlinks=true, allcolors=blue]{hyperref}

\usepackage{algorithm, algorithmic}

\newtheorem{theorem}{Theorem}

\providecommand{\keywords}[1]
{ \small \textbf{Keywords:} #1 }

\title{A joint estimation approach for monotonic regression functions in general dimensions}
\author{Christian Rohrbeck$^1$\footnote{Corresponding author. Email: 
\href{mailto:cr777@bath.ac.uk}{cr777@bath.ac.uk}}~ and Deborah A.~Costain$^2$}
\date{$^1$ Department of Mathematical Sciences, University of Bath\\ $^2$ Department of Mathematics and Statistics, Lancaster University}

\begin{document}

\maketitle
\doublespacing

\begin{abstract}
Regression analysis under the assumption of monotonicity is a well-studied statistical problem and has been used in a wide range of applications. However, there remains a lack of a broadly applicable methodology that permits information borrowing, for efficiency gains, when jointly estimating multiple monotonic regression functions. We introduce such a methodology by extending the isotonic regression problem presented in the article \textit{The isotonic regression problem and its dual} \citep{BarlowBrunk1972}. The presented approach can be applied to both fixed and random designs and any number of explanatory variables (regressors). Our framework penalizes pairwise differences in the values (levels) of the monotonic function estimates, with the weight of penalty being determined based on a statistical test, which results in information being shared across data sets if similarities in the regression functions exist. Function estimates are subsequently derived using an iterative optimization routine that uses existing solution algorithms for the isotonic regression problem. Simulation studies for normally and binomially distributed response data illustrate that function estimates are consistently improved if similarities between functions exist, and are not oversmoothed otherwise. We further apply our methodology to analyse two public health data sets: neonatal mortality data for Porto Alegre, Brazil, and stroke patient data for North West England.
\end{abstract}

\section{Introduction}

When analyzing the relationship between a univariate response $Y$ and a set of explanatory variables $\mathbf{X}$, we often have prior knowledge on the shape of the regression function~$\lambda$, which describes the expectation of $Y$ conditional on $\mathbf{X}$: $\mathbb{E}(Y\mid\mathbf{X}=\mathbf{x})=\lambda(\mathbf{x})$. For instance, dose-response curves and economic demand functions are usually monotonic \citep{Royston2000, Shively2009, Wilson2014}, and models for option pricing and health care planning and provision may impose convexity \citep{Ait2003, Hannah2013}. In such applications, incorporating shape constraints in the inference framework can have a substantial stabilizing effect, and prevent overfitting.

Regression analysis subject to the constraint of $\lambda$ being monotonic, an assumption that can be readily tested \citep{Bowman1998, Scott2015}, is the most widely studied case of shape-constraint inference, and is referred to as isotonic regression. This has resulted in the development of a variety of methods for function estimation under monotonicity constraints \citep{BarlowBrunk1972, Bacchetti1989, Gelfand1991, Dunson2005, Saarela2011}, some of which impose additional restrictions on $\lambda$, such as continuity \citep{Ramsay1998, Shively2009, Lin2014}. A number of these methods and their theoretical properties are reviewed later in Section~\ref{sec:Review}.

We consider the statistical problem of jointly estimating $K$ monotonic functions~$\lambda_1,\ldots,\lambda_K$, where $\lambda_k~(k=1,\ldots,K)$ describes the expectation of a response~$Y_k$ conditional on the vector  $\mathbf{X}_k$ of explanatory variables, $\mathbb{E}(Y_k\mid\mathbf{X}_k= \mathbf{x})= \lambda_k(\mathbf{x})$. The following example from personalized medicine illustrates the motivation for our work: To account for dose-response curves being dependent on  patients' characteristics, patients with similar characteristics are assigned to one of the $K$~experimental groups, and data are collected for each group. While the dose-response curves $\lambda_1,\ldots,\lambda_K$ can then be estimated separately for each group, more efficient estimates may be obtained by borrowing information across groups which exhibit a similar dose-response relationship at certain doses. This, however, requires us to carefully decide between which groups statistical information is to be shared, and to which extent.  

There exists surprisingly little research on the joint estimation of multiple monotonic functions.  \citet{Sasabuchi1983} and \citet{Nomakuchi1988} formulate a multivariate analogue to \citet{BarlowBrunk1972}. However, they do not exploit potential similarities in $\lambda_1,\ldots,\lambda_K$ for efficiency gains, but instead consider the case of $Y_k\mid(\mathbf{X}_k=\mathbf{x}_k)$ and $Y_p\mid(\mathbf{X}_p=\mathbf{x}_p)$ ~($k\neq p$) being correlated with known covariance structure. \citet{Gelfand1991} and \citet{Saarela2020} apply the monotonicity assumption in an ordinal regression framework with layered / stacked functions. Their approaches only borrow information in so far as to preserve the ordering amongst the survival curves representing the different categories, which is quite different from the set up considered herein. \citet{Rohrbeck2017} are the first to borrow statistical information for the joint estimation of multiple monotonic functions for efficiency gains, but their approach scales poorly with increasing $K$.

This paper introduces an inference framework which exploits similarities in the shapes of some, or all, of the functions $\lambda_1,\ldots,\lambda_K$ to increase estimation efficiency in both fixed and random designs. Our approach does not require prior knowledge, or assumptions, on which functions have similar shapes. Instead, we decide on the imposition and strength of the borrowing across any pair ($\lambda_k,\lambda_p$)~($k\neq p$) of monotonic functions at a point $\mathbf{x}$ by applying a statistical test on whether $\lambda_k(\mathbf{x})$ and $\lambda_p(\mathbf{x})$ are identical; properties of the proposed test are studied using simulated data. We incorporate this information within a convex optimization problem, generalising the isotonic regression problem by \citet{BarlowBrunk1972}. The function estimates for $\lambda_1,\ldots,\lambda_K$ are then obtained by iteratively applying existing solution algorithms for the isotonic regression problem in order to solve the defined convex optimization problem. Our methodology provides two key advantages over \citet{Rohrbeck2017}: (i)~increased computational efficiency and (ii) no prior knowledge on the similarities of the functions is required.  

The statistical problem considered herein is in part linked to transfer learning, where interest lies in the estimation of the distribution of a target population, while also having access to data from a related population. Recent publications in transfer learning, such as \citet{Cai2021} and \citet{Reeve2021}, consider the task of deciding whether the related data should be used, together with the data for the target population, for estimation or not. In our case we have to similarly decide whether there is sufficient evidence to conclude that the function estimate for $\lambda_k$~($k=1,\ldots,K$) is likely to improve if we use the data for $\lambda_1,\ldots,\lambda_{k-1},\lambda_{k+1},\ldots,\lambda_K$.

The remainder of this paper is organized as follows: Section~\ref{sec:Review} provides a review of the existing literature on estimating a single monotonic function using optimization. Section~\ref{sec:Method} details our inference procedure for fixed and random designs and Section~\ref{sec:Sim} illustrates performance of our methodology using simulated data. Two real-world applications are considered in Section~\ref{sec:CaseStudy} and we conclude with a discussion in Section~\ref{sec:Discussion}.

\section{Background}
\label{sec:Review}

We wish to explore the association between a univariate response $Y$ and the $m$ explanatory variables in~$\mathbf{X}=(X_1,\ldots,X_m)$. In most isotonic regression settings, the relationship between $Y$ and $\mathbf{X}$ is assumed to be
\begin{equation}
Y~=~\lambda\left(\mathbf{X}\right)~+~\epsilon,
\label{eq:IsoRegProb}
\end{equation}
where $\lambda:\mathbb{R}^m\to\mathbb{R}$ is order-preserving (isotonic) and $\epsilon$ is a random variable with mean zero and finite variance, i.e., $\mathbb{E}\left(Y\mid\mathbf{X}=\mathbf{x}\right)=\lambda(\mathbf{x})$. Given an ordering $\preceq$ on $\mathbb{R}^m$, a function~$\lambda$ is isotonic if for any $\mathbf{u},\mathbf{v}\in\mathbb{R}^m$, with $\mathbf{u}\preceq\mathbf{v}$, $\lambda(\mathbf{u})\leq\lambda(\mathbf{v})$.  Note that any monotonic function can be made isotonic by reversing some of the coordinate axes. Throughout this paper, we take~$\preceq$ to be the Euclidian ordering, but alternative orderings and definitions of multivariable monotonicity exist, for instance, \citet{Fang2021}.

Let $\mathcal{D}=\left\{\,\left(y_i,\mathbf{x}_i\right)\in\mathbb{R}\times\mathbb{R}^m\,:\,i=1,\ldots,n\,\right\}$ denote the set of observations for $Y$ and $\mathbf{X}$. The isotonic regression problem by \citet{BarlowBrunk1972} defines the estimates for $\lambda\left(\mathbf{x}_1\right),\ldots,\lambda\left(\mathbf{x}_n\right)$ as the values $\hat{y}_{1},\ldots,\hat{y}_{n}$ which minimize the objective function
\begin{equation}
\sum_{i=1}^{n}\, w_i\left(\,y_i \, - \, \hat{y}_i\,\right)^2,
\label{eq:BarlowBrunk1972}
\end{equation}
subject to $w_1\ldots,w_n>0$ being fixed, and the monotonicity constraint that $\hat{y}_i\leq\hat{y}_j$ when $\mathbf{x}_i\preceq \mathbf{x}_j$ $\left(i,j=1,\ldots,n\right)$. Solution algorithms are, for instance, presented by \citet{Dykstra1982} and \citet{Luss2012}. An estimate $\hat{\lambda}$ of $\lambda$ is then obtained via interpolation; for instance, defining 
\begin{equation}
\hat{\lambda}(\mathbf{z})~=~\max_{i=1,\ldots,n}\, \left\{\, \hat{y}_i~:~\mathbf{x}_i\preceq \mathbf{z}\, \right\},\qquad \mathbf{z}\in\mathbb{R}^m, 
\label{eq:Interpolation}
\end{equation}
yields $\hat{\lambda}$ being isotonic and piecewise constant with $\hat\lambda\left(\mathbf{x}_i\right) = \hat{y}_i$~($i=1,\ldots,n$); other interpolation techniques, such as splines, may be applied to satisfy various smoothness assumptions on $\lambda$. A variety of extensions to the isotonic regression problem have been proposed: \citet{Bacchetti1989} incorporates the additional assumption that $\lambda$ has an additive form; \citet{Luss2014} replace the squared difference in the objective function \eqref{eq:BarlowBrunk1972} by a convex loss function; \citet{Sasabuchi1983} consider a multivariate analogue to~\eqref{eq:IsoRegProb} with correlated responses. 

Properties of the estimators $\hat{Y}_1,\ldots,\hat{Y}_n$, defined as the solutions to the isotonic regression problem, have been widely studied. \citet{Hanson1973} and \citet{Robertson1975} examine pointwise convergence of $\hat{Y}_i$~($i=1,\ldots,n$) and find $\hat{Y}_i \overset{p}{\to} \lambda(\mathbf{x}_i)$ as $n\to\infty$ for continuous~$\lambda$ and under certain conditions on the distribution of $\mathbf{X}$. For settings with a single explanatory variable, $m=1$, \cite{Wright1981}, amongst others, show that the estimator $\hat{Y}_{i}$ converges to the true value~$\lambda(x_i)$ at a rate of $n^{-1/3}$ for sufficiently smooth $\lambda$, with its asymptotic behaviour being described by Chernoff's distribution. Further, again for $m=1$, \citet{Meyer2000} provide an upper bound on the expected number of distinct values of $\hat{Y}_{1},\ldots,\hat{Y}_{n}$ for $\epsilon\sim\mbox{Normal}(0,\sigma^2)$ in~\eqref{eq:IsoRegProb}. Note, the sampling variance of $\hat{Y_i}$ and the expected number of distinct values both increase with the gradient of $\lambda$. Properties of the vector $(\hat{Y}_1,\ldots,\hat{Y}_n)$ are, for instance, explored by \citet{Zhang2002}, \citet{Chatterjee2018} and \citet{Han2019}, with the latter, 2019, authors being the only ones to consider the case of $\mathbf{X}$ being randomly distributed. For $\lambda:[0,1]\to\mathbb{R}$ smooth, global convergence of the estimator for $\lambda$, based on the interpolation~\eqref{eq:Interpolation}, has been studied by \citet{Durot2002}.

\section{Methodology}
\label{sec:Method}

We consider a setting with $K$~unknown monotonic functions $\lambda_1,\ldots,\lambda_K$, where $\lambda_k$~($k=1,\ldots,K$), similar to \eqref{eq:IsoRegProb}, describes the relationship between the response $Y_k$ and the $m$ explanatory variables in $\mathbf{X}_k$,
\begin{equation}
Y_k = \lambda_k(\mathbf{X}_k) + \epsilon_k.
\label{eq:GeneralIsoReg}
\end{equation}
The random variables $\epsilon_1,\ldots,\epsilon_K$ are assumed to be independently distributed with mean zero and finite (but potentially different) variance. 

This section introduces an inference framework for $\lambda_1,\ldots,\lambda_K$ which exploits similarities across pairs of functions for efficiency gains, in fixed and random designs for $\mathbf{X}_k$, and does not require any prior assumptions on which functions are similar. Application of our method only requires the specification of a single parameter $\alpha\in(0,1)$, with the potential efficiency gains and the risk of oversmoothing both increasing with decreasing $\alpha$.

Section~\ref{sec:SimilarityMeasure} considers testing for equivalence of $\lambda_k(\mathbf{x})$ and $\lambda_p(\mathbf{x})$ ($k\neq p$) for a given $\mathbf{x}\in\mathbb{R}^m$ in a fixed design with no missing values. A multivariate isotonic regression problem that permits joint estimation of $\lambda_1,\ldots,\lambda_K$ is introduced in Section~\ref{sec:OptimizationProblem}, and a solution algorithm is presented in Section~\ref{sec:Estimation}. The generalization to missing data and random designs is described in Section~\ref{sec:RandomDesign}.

\subsection{Assessing point-wise similarity of monotonic function pairs}
\label{sec:SimilarityMeasure}

Let $\lambda_k$ and $\lambda_p$ ($k\neq p$) be two monotonic functions within isotonic regression models of the form~\eqref{eq:GeneralIsoReg}. The observations for $(Y_k,\mathbf{X}_k)$ and $(Y_p,\mathbf{X}_p)$ are denoted by $\mathcal{D}_k=\{(y_{k,i},\mathbf{x}_{k,i}):i=1,\ldots,n\}$ and $\mathcal{D}_p=\{(y_{p,i},\mathbf{x}_{p,i}):i=1,\ldots,n\}$ respectively, where $\mathbf{x}_{k,i}=\mathbf{x}_{p,i}$ for all $i=1,\ldots,n$. Without loss of generality, the functions are assumed to be isotonic and to operate on a $m$-dimensional hypercube, such as $\lambda_k,\lambda_p:[0,1]^m\to\mathbb{R}$. The indices $k$ and $p$ in $\mathbf{x}_{k,i}$ and  $\mathbf{x}_{p,i}$ are dropped for brevity in the following. 

For design point $\mathbf{x}_t$~($t=1,\ldots,n$), we wish to determine whether there is sufficient evidence to conclude that $\lambda_k(\mathbf{x}_t) \neq \lambda_p(\mathbf{x}_t)$. This can be expressed as a statistical test with null hypothesis $H_0:\lambda_k(\mathbf{x}_t)=\lambda_p(\mathbf{x}_t)$ and alternative hypothesis $H_1:\lambda_k(\mathbf{x}_t)\neq \lambda_p(\mathbf{x}_t)$. Such a test can, in general, not be performed by only considering the data points associated with~$\mathbf{x}_t$, because function estimates under the null and alternative hypothesis have to satisfy the monotonicity constraints. Moreover, in many applications we only have a single observation at $\mathbf{x}_t$ which yields a large Type~II error. 

Under the alternative hypothesis, the estimates providing the best model fit correspond to the values $\hat{y}_{k,1},\ldots,\hat{y}_{k,n}$ and $\hat{y}_{p,1},\ldots,\hat{y}_{p,n}$ obtained by solving the isotonic regression problem in Section~\ref{sec:Review} for the data sets $\mathcal{D}_k$ and $\mathcal{D}_p$ separately. With the additional constraint of the function estimates for $\lambda_k$ and $\lambda_p$ being identical at $\mathbf{x}_t$, the estimates $\hat{y}_{k,1}^{(0)},\ldots,\hat{y}_{k,n}^{(0)}$ and $\hat{y}_{p,1}^{(0)},\ldots,\hat{y}_{p,n}^{(0)}$ under the null hypothesis minimize the objective function
\begin{equation}
\sum_{i=1}^n w_{k,i} \left(y_{k,i}-\hat{y}_{k,i}^{(0)}\right)^2 + \sum_{i=1}^n w_{p,i} \left(y_{p,i}-\hat{y}^{(0)}_{p,i}\right)^2,
\label{eq:TestOptim}
\end{equation}
subject to the constraints $\hat{y}_{k,i}^{(0)}\leq\hat{y}_{k,j}^{(0)}$ and $\hat{y}_{p,i}^{(0)}\leq\hat{y}_{p,j}^{(0)}$, if $\mathbf{x}_i\preceq \mathbf{x}_j$ $\left(i,j=1,\ldots,n\right)$, and $\hat{y}_{k,t}^{(0)}=\hat{y}_{p,t}^{(0)}$. This optimization problem corresponds to an isotonic regression problem with $2n-1$ parameters, and we solve it using existing solution algorithms. Given the estimates derived under the null and alternative hypothesis, we consider a likelihood ratio test with test statistic
\begin{equation}
\mathrm{LR}(k,p,t)= 
-2 \left\{ 
\ell\left(\mathbf{y}_k\mid\hat{\mathbf{y}}_k^{(0)}, \bm\theta_k\right) +
\ell\left(\mathbf{y}_p\mid\hat{\mathbf{y}}_p^{(0)}, \bm\theta_p\right) -
\ell\left(\mathbf{y}_k\mid\hat{\mathbf{y}}_k, \bm\theta_k\right) -
\ell\left(\mathbf{y}_p\mid\hat{\mathbf{y}}_p, \bm\theta_p\right)
\right\},
\label{eq:LR}
\end{equation}
where $\ell(\cdot)$ is the log-likelihood function for the model in \eqref{eq:GeneralIsoReg} and $\bm\theta_k$ and $\bm\theta_p$ are the known (or estimated) parameters describing the distributions of $\epsilon_k$ and $\epsilon_p$ respectively. 

It remains for us to determine the critical value to compare $\mathrm{LR}(k,p,t)$ to when deciding whether to reject $H_0$ or not. A likelihood ratio test would suggest comparing $\mathrm{LR}(k,p,t)$ to the quantile of a chi-squared distribution. However, the difference in dimensions between the two models cannot be easily determined since we work within a semi-parametric modelling framework. Specifically, the additional constraint $\hat{y}_{k,t}^{(0)}=\hat{y}_{p,t}^{(0)}$ is not equivalent to the restricted model having one dimension less than the alternative model.

\cite{Meyer2000} argue that the effective number of dimensions in the isotonic regression problem \eqref{eq:BarlowBrunk1972} corresponds to the number of distinct values among the estimates $\hat{y}_{k,1},\ldots,\hat{y}_{k,n}$. We thus added to the optimization problem with objective function \eqref{eq:TestOptim} the constraint that the number of distinct values combined across $\hat{y}^{(0)}_{k,1},\ldots,\hat{y}^{(0)}_{k,n}, \hat{y}^{(0)}_{p,1},\ldots,\hat{y}^{(0)}_{p,n}$ had to be smaller than that of $\hat{y}_{k,1},\ldots,\hat{y}_{k,n}, \hat{y}_{p,1},\ldots,\hat{y}_{p,n}$. This should imply that the effective dimension under the null hypothesis is smaller than that under the alternative. However, this approach only led to minor changes in the simulated quantiles of the test statistic and we thus proceed without this additional constraint in the rest of the paper.

To explore the distribution of the test statistic $\mathrm{LR}(k,p,t)$, we ran a simulation study with $m=1$, $\lambda_k(x)=\lambda_p(x)=x^2$, $x\in[0,4]$, and $\epsilon_k,\epsilon_p\sim\mathrm{Normal}(0,1)$. The fixed design points were set to $x_i=4i/n$~($i=1,\ldots,n$) with $n\in\{50,200,1000,2000\}$, and the test statistic was calculated at the values $x=1,2,3$ for 20,000 generated data sets. This set up facilitates the investigation of the impact of the gradient of $\lambda_k$ on the distribution of the test statistic; several authors, for instance \cite{Wright1981}, have shown that the gradient affects the rate of convergence of the estimate $\hat{y}_{k,i}$ to the true value $\lambda_k(x_{i})$~$(i=1,\ldots,n)$ as $n\to\infty$.

\begin{table}
\centering
\begin{tabular}{lc|ccccc}
\hline
Sample size& Value~~~&$q_{0.5}$&$q_{0.7}$&$q_{0.8}$&$q_{0.9}$& $q_{0.95}$\\ 
\hline
         & $x=1$ &0.306 &0.718 &1.080 &1.777 &2.512\\
$n=50$   & $x=2$ &0.342 &0.795 &1.195 &1.945 &2.730\\
         & $x=3$ &0.348 &0.822 &1.244 &2.057 &2.887\\
\hline         
         & $x=1$ &0.287 &0.665 &1.001 &1.625 &2.236\\
$n=200$  & $x=2$ &0.299 &0.709 &1.076 &1.738 &2.400\\
         & $x=3$ &0.310 &0.712 &1.090 &1.766 &2.459\\
\hline         
         & $x=1$ &0.278 &0.633 &0.965 &1.565 &2.198\\
$n=1000$ & $x=2$ &0.276 &0.638 &0.966 &1.586 &2.244\\
         & $x=3$ &0.276 &0.653 &1.010 &1.633 &2.305\\
\hline        
         & $x=1$ &0.263 &0.614 &0.929 &1.528 &2.162\\
$n=2000$ & $x=2$ &0.272 &0.640 &0.968 &1.580 &2.211\\
         & $x=3$ &0.274 &0.636 &0.972 &1.608 &2.250\\
\hline         
\end{tabular}
\caption{Quantiles of $\mathrm{LR}(k,p,t)$ based on $20,000$ samples with $\lambda_k(x)=\lambda_p(x)=x^2$, sample size $n=\{50,200,1000,2000\}$ and $x=1,2,3$.}
\label{tab:DistirbutionLR}
\end{table}

Table~\ref{tab:DistirbutionLR} indicates that the gradient has an affect on the distribution of $\mathrm{LR}(k,p,t)$, with the quantiles being slightly higher for the value $x=3$, the point at which $\lambda_k$ has the highest gradient amongst the considered values for $x$. Further, the differences between the quantiles of $\mathrm{LR}(k,p,t)$ for $x=1$, $x=2$ and $x=3$ decrease with increasing sample size $n$, for instance, the 95\% quantiles at $x=1$ and $x=3$ differ by 0.35 for $n=50$ and by $0.10$ for $n=1000$. We also studied settings with $\epsilon_k~(\epsilon_p)$ having higher variance and functions with a smaller range of values than for $\lambda_1(x)=\lambda_2(x)=x^2$. The conclusions drawn from these simulations were similar to that in Table~\ref{tab:DistirbutionLR}, with a higher variance resulting in slightly lower quantile estimates. While the results do not show whether the asymptotic distribution of $\mathrm{LR}(k,p,t)$ is degenerate, the estimated quantiles suggest that the upper quantiles of the distribution of $\mathrm{LR}(k,p,t)$ are smaller than that of a chi-squared distribution with one degree of freedom. Consequently, deriving the critical value from a chi-squared distribution with one degree of freedom results in a conservative test, which may be preferred in some situations. We exploit this aspect in the next Section~\ref{sec:OptimizationProblem} when deciding whether to borrow statistical information across the observations in $\mathcal{D}_k$ and $\mathcal{D}_p$. 

To investigate performance of the testing procedure when using the chi-squared distribution with one degree of freedom, we considered a two function setting with $\lambda_1(x)=x^2$, $\lambda_2(x) = x^2 \mathbbm{1}\{x\leq 2\} + (x+2)\mathbbm{1}\{x>2\}$ and $n=200$ fixed design points, where $x_i=4i/n$. For $x\in[0,2]$, the points at which the functions are identical (i.e.~$H_0$ is true), the average rejection rate at the 5\% significance level was 2.1\% and for $x>2.5$ ($H_1$ true) we almost always rejected $H_0$; see Section A in the online supplementary materials for details.

\subsection{A multivariate isotonic regression problem}
\label{sec:OptimizationProblem}

Let's consider the task of estimating the functions $\lambda_1,\ldots,\lambda_K$. If we assume no similarity across the function set, the $K$ objective functions of the form \eqref{eq:BarlowBrunk1972} can be combined to 
\[
\sum_{k=1}^K \sum_{i=1}^n w_{k,i} \left(y_{k,i}-\hat{y}_{k,i}\right)^2
\]
with the monotonicity constraints $\hat{y}_{k,i}\leq \hat{y}_{k,j}$ for all $k=1,\ldots,K$ when $\mathbf{x}_i\preceq \mathbf{x}_j$~($i,j=1,\ldots,n$). To borrow information across the data sets for $\lambda_1,\ldots,\lambda_K$, we add a second component to the objective function which penalizes pairwise differences of the fitted function levels,
\begin{equation}
\sum_{k=1}^K\, \sum_{i=1}^n \left\{
w_{k,i}\left(\,y_{k,i}\,-\,\hat{y}_{k,i}\,\right)^2\, +\,
\frac{1}{2}\sum_{p\neq k}\, v_{k,p,i} \left(\,\hat{y}_{k,i}\,-\,\hat{y}_{p,i}\,\right)^2
\right\},
\label{eq:OptimExtended}
\end{equation}
where $v_{k,p,i}\geq 0$ refers to the weight given to the squared difference of $\hat{y}_{k,i}$ and $\hat{y}_{p,i}$: higher values for $v_{k,p,i}$ lead to $\hat{y}_{k,i}$ and $\hat{y}_{p,i}$ being more similar. Our defined optimization problem shares an important property with the isotonic regression problem by \cite{BarlowBrunk1972}:

\begin{theorem}
\label{Theorem1}
For $w_{k,i}>0$ and $v_{k,p,i}\geq 0$~($i=1,\ldots,n;~k,p=1,\ldots,K$), there exists a unique solution to the optimization problem of minimizing the objective function \eqref{eq:OptimExtended} subject to $\hat{y}_{k,i}\leq \hat{y}_{k,j}$ for all $k=1,\ldots,K$ when $\mathbf{x}_i\preceq \mathbf{x}_j$.
\end{theorem}

\begin{proof}
To prove uniqueness, it is sufficient to show that the defined optimization problem is convex, that is, both the objective function and the set of feasible solutions are convex. Firstly, the objective function \eqref{eq:OptimExtended} is strictly convex because it is the sum of strictly convex functions. Secondly, \citet{BarlowBrunk1972} note that the constraints on $\hat{y}_1,\ldots,\hat{y}_n$ in~\eqref{eq:BarlowBrunk1972} define a convex cone in $\mathbb{R}^n$. Further, we impose no constraints between function estimates, for instance, $\hat{y}_{k,i}$ does not affect the set of feasible solutions for $\hat{y}_{p,i}$ $(k\neq p)$. Consequently, the set of feasible solutions to the optimization problem with objective function \eqref{eq:OptimExtended} corresponds to the Cartesian product of $K$~convex cones and, as such, is a convex set.  
\end{proof}

Our proposal~\eqref{eq:OptimExtended} is not too dissimilar in form to the objective function used in ridge regression, highlighting an interesting connection to approaches on model sparsity and variable selection. Therefore, we may, for instance, replace the squared difference by a lasso type penalty, leading to the Lagrangian form of a fused lasso, with the additional monotonicity constraints. However, solving such an optimization problem is more challenging than minimizing the objective function \eqref{eq:OptimExtended} subject to monotonicity constraints. Furthermore, our main interest lies in increasing model robustness to outliers (through borrowing), instead of imposing that function estimates are identical at a certain point in $\mathbb{R}^m$ if they appear sufficiently similar.

We are left with specifying the set $\{v_{k,p,i}:k,p=1,\ldots,K;~i=1,\ldots,n\}$ of smoothing weights. Our proposal is to define
\begin{equation}
v_{k,p,i} = \max\left\{0~,~1 - \frac{\mathrm{LR}(k,p,i)}{F_Q^{-1}(1-\alpha)}\right\},
\label{eq:weight}
\end{equation}
where $F_Q^{-1}(\cdot)$ is the quantile function of a chi-squared distribution with one degree of freedom and $\alpha\in(0,1)$ has to be specified by the user. This formulation leads to no smoothing being applied, $v_{k,p,i}=0$, when the null hypothesis in Section~\ref{sec:SimilarityMeasure} is rejected at the $(100\times\alpha)\%$ significance level, while $v_{k,p,i}>0$ otherwise. Consequently, the degree of smoothing across function estimates is determined by the level of oversmoothing we are willing to tolerate in the form of the specified parameter $\alpha$; smaller values of $\alpha$ yield a higher risk of oversmoothing, but also an increased potential for efficiency gains. 

As discussed in Section~\ref{sec:SimilarityMeasure}, applying the chi-squared distribution with one degree of freedom results in a conservative test. Hence we borrow information at points where functions pairs are identical with probability above $(1-\alpha)$, but this may result in oversmoothing at the points where the functions are different. We perform simulations in Section~\ref{sec:Sim} to investigate sensitivity to the choice of $\alpha$.

\subsection{Solution algorithm and performance measures}
\label{sec:Estimation}

We employ a cyclical optimization procedure to find the solution to the optimization problem defined in Section~\ref{sec:OptimizationProblem}; a similar strategy has been used by \citet{Sasabuchi1983} and \citet{Bacchetti1989} for other extensions of the isotonic regression problem. The initial solution is set to the estimates for the isotonic regression problem by \citet{BarlowBrunk1972}. At each iteration, we minimize~\eqref{eq:OptimExtended} with respect to $\mathbf{\hat{y}}_k=(\hat{y}_{k,1},\ldots,\hat{y}_{k,n})$ ($k=1,\ldots,K$) while keeping the estimates for the remaining $K-1$ functions fixed. Functions are updated in-turn until convergence; this is equivalent to a block coordinate descent algorithm. The following Algorithm \ref{SolutionAlgorithm} provides a sketch of our full approach.

\begin{algorithm}
\caption{Solution algorithm for the optimization problem \eqref{eq:OptimExtended}}
\label{SolutionAlgorithm}
\begin{algorithmic}[1]
\REQUIRE Data sets $\mathcal{D}_1,\ldots,\mathcal{D}_K$
\REQUIRE Constants $w_{1,1},\ldots,w_{K,n}$ in \eqref{eq:BarlowBrunk1972}
\STATE Derive initial estimates for $\mathbf{\hat{y}}_1,\ldots,\mathbf{\hat{y}}_K$ using \cite{BarlowBrunk1972}
\STATE Calculate the smoothing constants $v_{1,2,1},\ldots,v_{K-1,K,n}$
\REPEAT
\FOR{k in 1 to K}
\STATE Update $\hat{\mathbf{y}}_k$ by minimizing \eqref{eq:OptimExtended}, with $\hat{\mathbf{y}}_1,\ldots,\hat{\mathbf{y}}_{k-1},\hat{\mathbf{y}}_{k+1},\ldots,\hat{\mathbf{y}}_K$ being fixed
\ENDFOR
\UNTIL $\hat{\mathbf{y}}_1,\ldots,\hat{\mathbf{y}}_K$ have converged
\ENSURE Optimal solution $\hat{\mathbf{y}}_1, \ldots, \hat{\mathbf{y}}_K$
\end{algorithmic}
\end{algorithm}

For $\hat{\mathbf{y}}_1,\ldots,\hat{\mathbf{y}}_{k-1},\hat{\mathbf{y}}_{k+1},\ldots, \hat{\mathbf{y}}_K$ fixed, solving the optimization problem with objective function \eqref{eq:OptimExtended} for $\mathbf{\hat{y}}_k$ is equivalent to minimizing
\begin{equation}
\sum_{i=1}^n \left(w_{k,i}+\sum_{p\neq k} v_{k,p,i}\right) \left( \hat{y}_{k,i} - \frac{w_{k,i} y_{k,i}+\sum_{p\neq k} v_{k,p,i} \hat{y}_{p,i}}{w_{k,i}+\sum_{p\neq k} v_{k,p,i}}\right)^2,
\label{eq:UpdateLambdak}
\end{equation}
subject to the monotonicity constraints. Therefore, the minimization in line 5 of Algorithm \ref{SolutionAlgorithm} can be written as an optimization problem of the form \eqref{eq:BarlowBrunk1972} and solved using existing solution algorithms for the isotonic regression problem. Herein, the R package \texttt{isotone} \citep{Leeuw2011} is used for cases with $m=1$ explanatory variables, and we implemented isotonic recursive partitioning \citep{Luss2012} to handle cases with $m>1$. 

Algorithm \ref{SolutionAlgorithm} converges to the optimal solution because (i) the algorithm starts from a feasible solution, (ii) the objective function is non-negative, strictly convex and twice differentiable and (iii) we have a boundary on the possible values for the values $\mathbf{\hat{y}}_1,\ldots,\mathbf{\hat{y}}_K$. In particular, the estimates will not lie outside the range of observed values for $y_{1,1}, \ldots,y_{K,n}$; see for instance \cite{Tseng2001} and references therein for results on block coordinate descent algorithms.

Performance of our approach will be measured using two metrics that compare the estimates $\mathbf{\hat{y}}_1,\ldots,\mathbf{\hat{y}}_K$ derived from \eqref{eq:OptimExtended} to the estimates $\mathbf{\hat{y}}_1^{(B)},\ldots,\mathbf{\hat{y}}_K^{(B)}$ obtained using the isotonic regression problem by \cite{BarlowBrunk1972}. For estimates $\mathbf{\hat{y}}_1,\ldots,\mathbf{\hat{y}}_K$, the Root Mean Squared Error (RMSE) is defined as
\[
\mathrm{RMSE}\left(\mathbf{\hat{y}}_1,\ldots,\mathbf{\hat{y}}_K\right) = \sqrt{\,\frac{1}{K}\sum_{k=1}^K\,\frac{1}{n} \sum_{i=1}^{n}\,\left\{\lambda_k(\mathbf{x}_i)-\hat{y}_{k,i})\,\right\}^2}.
\]
The efficiency gain is then quantified via the ratio 
\[
R:=\frac{\mathrm{RMSE}\left(\mathbf{\hat{y}}_1,\ldots,\mathbf{\hat{y}}_K\right)} {\mathrm{RMSE}\left(\mathbf{\hat{y}}_1^{(B)},\ldots,\mathbf{\hat{y}}_K^{(B)}\right)},
\]
where $R<1$ corresponds to our approach performing better than \cite{BarlowBrunk1972} - we will derive the median and interquartile range of $R$ across multiple simulated data sets. To measure whether the achieved efficiency gains are robust (consistent), we derive an estimate $P$ for $\mathbb{P}(R<1)$, the probability that our method provides a lower RMSE than a method that applies no information borrowing. Values of $P$ close to 1 would indicate that we consistently improve estimates.

\subsection{Incomplete data and random designs}
\label{sec:RandomDesign} 

We so far assumed that the observations for $\mathbf{X}_k$ are fixed design points and that a realization of $Y_k\mid(\mathbf{X}_k=\mathbf{x})$ is observed for each $k=1,\ldots,K$ and all grid points $\mathbf{x}_1,\ldots,\mathbf{x}_n$. In practice this assumption may not hold, for instance, because none of the patients in a group received a dose that was given to members of a different experimental group. The key message is that our test in Section~\ref{sec:SimilarityMeasure} can be applied to any point $\mathbf{x}\in\mathbb{R}^m$. This allows us to again derive weights as in \eqref{eq:weight} and to define an optimization problem similar to that in Section~\ref{sec:OptimizationProblem} which can be applied to fixed designs with missing values and random designs.

Let $\mathbf{x}_t$ denote a combination of explanatory variables for which we observe $Y_k\mid(\mathbf{X}_k=\mathbf{x}_t)$ but not $Y_p\mid (\mathbf{X}_p=\mathbf{x}_t)$~($k\neq p$). As in Section~\ref{sec:SimilarityMeasure}, we wish to test $H_0:\lambda_k(\mathbf{x}_t)=\lambda_p(\mathbf{x}_t)$ against $H_1:\lambda_k(\mathbf{x}_t)\neq \lambda_p(\mathbf{x}_t)$. Under the null hypothesis, we again minimize the objective function \eqref{eq:TestOptim} subject to the monotonicity constraints and $\hat{y}_{k,t}^{(0)} = \hat{y}_{p,t}^{(0)}$ - the only difference is that $w_{p,t}=0$ since we have no observation for $Y_p\mid (\mathbf{X}_p=\mathbf{x}_t)$. The empirical quantiles of the test statistic $\mathrm{LR}(k,p,t)$ for the set up in Table~\ref{tab:DistirbutionLR} are provided in Section B of the online supplementary materials and indicate moderate differences (minor differences) for moderate (large) sample sizes.

While the test statistic $\mathrm{LR}(k,p,t)$ in \eqref{eq:LR} can be computed as before, some care has to be taken regarding the conversion of $\mathrm{LR}(k,p,t)$ to $v_{k,p,t}$ via \eqref{eq:weight}. One new aspect is that we may obtain $\mathrm{LR}(k,p,t)=0$, i.e.~$v_{k,p,t}=1$, despite the estimates $\hat{y}_{k,t}$ and $\hat{y}_{p,t}$ being different under the alternative hypothesis - this may occur when the constraint $\hat{y}_{k,t}^{(0)} = \hat{y}_{p,t}^{(0)}$ does not change the estimates at the observed data points in $\mathcal{D}_k$ and $\mathcal{D}_p$. In our numerical examples and case studies we set $v_{k,p,t}=1$ in this case, because there were no noticeable differences in performance - this approach may have to be considered more carefully in settings with a high number $K$ of functions. One alternative is to set $v_{k,p,t}=0$, but this becomes impractical in settings with random designs.

Let $n_k$ denote the number of points in $\mathcal{D}_k$~($k=1,\ldots,K$). Define $\mathcal{X}=\{\mathbf{x}_i:i=1,\ldots,n\}$ as the set of distinct observations for the explanatory variables across the $K$ groups; in a fixed design with no missing values $n=n_1=\cdots=n_K$, while $n\leq n_1+\cdots+n_K$ in random designs, where the upper bound corresponds to the observations of $\mathbf{X}_k$ being completely different to that for $\mathbf{X}_p$~$(k\neq p)$; see Section~\ref{sec:SimRandom} for an example. We further set $w_{k,i}=0$ and $v_{k,p,i}=0$ when $\mathbf{x}_{k,i}$~($i=1,\ldots,n$) is not contained in $\mathcal{D}_k$ and $\mathcal{D}_p$ respectively, and define the index set $\mathcal{I}_k=\{i:w_{k,i}>0\}\cup\bigcup_{p\neq k}\{i:v_{k,p,i}>0\}$. For a given set of weights, $\{v_{k,p,i}\,:\,k,p=1,\ldots,K;\, i=1,\ldots,n\}$, the estimates $\hat{\mathbf{y}}_k$ in Algorithm~\ref{SolutionAlgorithm}, for $\hat{\mathbf{y}}_1,\ldots,\hat{\mathbf{y}}_{k-1},\hat{\mathbf{y}}_{k+1},\ldots,\hat{\mathbf{y}}_K$ fixed, are updated by minimizing the objective function
\begin{equation}
\sum_{i\in\mathcal{I}_k} \left\{w_{k,i}\left(y_{k,i}-\hat{y}_{k,i}\right)^2 + \sum_{p\neq k} v_{k,p,i} \left(\hat{y}_{k,i} - \hat{y}_{p,i}\right)^2\right\},
\label{eq:UpdatekGeneral}
\end{equation}
subject to the monotonicity constraints. We can again change the objective function in the optimization problem to the form \eqref{eq:UpdateLambdak} and use existing solution algorithms for the isotonic regression problem. Note, we only derive estimates for $\hat{y}_{k,i}$ in \eqref{eq:UpdatekGeneral} if $w_{k,i}>0$ or $v_{k,p,i}>0$ for at least one~$p$. The estimates $\hat{\mathbf{y}}_1,\ldots,\hat{\mathbf{y}}_K$ are again updated in-turn until convergence as in Algorithm~\ref{SolutionAlgorithm}.

Our set up allows for information from other groups to be used to extrapolate the estimates related to $\lambda_k$ beyond the range of observed values in the data set $\mathcal{D}_k$. As for most flexible regression models, such an extrapolation relies on very strong assumptions and tends to be highly unreliable. In our simulations and case studies we decided against extrapolation and set $v_{k,p,i}=0$ whenever $\mathbf{x}_{i} \preceq \mathbf{x}_{k,j}~(j=1,\ldots,n_k)$ for all or none of the observations in $\mathcal{D}_k$. In the case $m=1$ this corresponds to $x_i$ lying to the left or to the right of the observations for the explanatory variable $X_k$.

\section{Numerical Examples}
\label{sec:Sim}

\subsection{Sensitivity analysis for \texorpdfstring{$\alpha$}{a}}
\label{sec:Sensitivity}

To explore sensitivity of our approach to the parameter $\alpha$ in \eqref{eq:weight}, we consider three studies with $K=2$ functions, $\lambda_1,\lambda_2:[0,4]\to\mathbb{R}$. In all studies, the observations for the explanatory variable are fixed design points that are equally spaced with $x_i=4i/n~(i=1,\ldots,n)$. The sample size is $n\in\{50,100,150,200\}$, and we consider $\alpha\in\{0.05,0.10\}$. Further, $\epsilon_1,\epsilon_2 \sim\mathrm{Normal}(0,\sigma^2)$ in \eqref{eq:GeneralIsoReg} with $\sigma^2=1$ known. For each considered setting, we generate 200 data sets and the results for the measures $R$ and $P$ (defined in Section~\ref{sec:Estimation}) are presented in Table~\ref{tab:Sensitivity}.

Study 1 considers $\lambda_1(x)=x^2$ and $\lambda_2(x)=x^2\mathbbm{1}\{x\leq2\}+(x+2)\mathbbm{1}\{x>2\}$, that is, the function pair is identical across the interval $[0,2]$ and exhibits substantially different levels for higher values of $x$. Since $P$ is close to 1 for all considered $n$ in Table~\ref{tab:Sensitivity}, our approach consistently outperforms the estimates obtained using the framework by \cite{BarlowBrunk1972}. The rate of efficiency gain is also consistent across the considered sample sizes, with the setting $\alpha=0.05$ performing slightly better than $\alpha=0.10$. 

\begin{table}
\centering
\begin{tabular}{cc|cc|cc}
\hline
Study & Sample size $n$ & \multicolumn{2}{|c|}{$\alpha=0.05$} & \multicolumn{2}{c}{$\alpha=0.10$} \\
  &  & $R$                 &  $P$  & $R$ & $P$\\
\hline
  & 50  &0.94 (0.92,0.96) &0.985 &0.95 (0.93,0.96) &0.980\\
1 & 100 &0.94 (0.92,0.96) &0.975 &0.95 (0.93,0.96) &0.995\\
  & 150 &0.94 (0.91,0.95) &1.000 &0.95 (0.93,0.96) &1.000\\
  & 200 &0.94 (0.92,0.95) &0.995 &0.95 (0.93,0.96) &1.000\\
\hline
  & 50  & 0.85 (0.79,0.90) &0.990 &0.89 (0.83,0.92) &0.990\\
2 & 100 & 0.88 (0.83,0.92) &0.995 &0.90 (0.86,0.93) &0.990\\
  & 150 & 0.90 (0.86,0.93) &0.975 &0.92 (0.88,0.94) &0.980\\
  & 200 & 0.91 (0.88,0.94) &0.965 &0.93 (0.90,0.96) &0.965\\
\hline
  & 50  & 1.01 (1.00,1.01) &0.320 &1.01 (1.00,1.01) &0.300\\
3 & 100 & 1.01 (1.00,1.01) &0.255 &1.00 (1.00,1.01) &0.290\\
  & 150 & 1.00 (1.00,1.01) &0.250 &1.00 (1.00,1.01) &0.265\\
\hline
\end{tabular}
\caption{Sensitivity of the performance measures $R$ and $P$ to sample size $n$ and significance level $\alpha$. The values for $R$ correspond to the median (inter-quartile range). The function pair in Study 1 is $\lambda_1(x)=x^2$ and $\lambda_2(x)=x^2\mathbbm{1}\{x\leq2\}+(x+2)\mathbbm{1}\{x>2\}$, while $\lambda_1(x)=1.95$ and $\lambda_2(x)=1 +\mathrm{expit}(x)$ in Study 2, and $\lambda_1(x)=x^2$ and $\lambda_2(x)=x^4-1$ in Study 3.}
\label{tab:Sensitivity}
\end{table}

In Study 2 the functions are $\lambda_1(x) = 1.95$ and $\lambda_2(x) = 1 + \mathrm{expit}(x)$, where $\mathrm{expit}(x)=\{1+\exp(-x)\}^{-1}$. Table~\ref{tab:Sensitivity} again shows a consistent improvement across all considered settings of $n$ and $\alpha$. Although $\lambda_1$ and $\lambda_2$ are only identical at a single point, their strong similarity for higher values of $x$ still invites borrowing to improve function estimates, in particular for small $n$. Note, these differences between $\lambda_1$ and $\lambda_2$ lead to $v_{k,p,i}=0$~($i=1,\ldots,n$) tending towards $0$ with increasing $n$, and thus to the small increase in $R$ for larger $n$.

Finally, we set $\lambda_1(x)=x^2$ and $\lambda_2(x) = x^4-1$ in Study 3 to investigate performance for a setting where functions cross at a point but are quite different otherwise. Compared to Study 2 we only have a very small interval across which the functions are similar, limiting the potential benefits of borrowing. In this case the interquartile range of $R$ in Table~\ref{tab:Sensitivity} is [1,1.01] for both $\alpha=0.05$ and $\alpha=0.10$ (results for $n=200$ not shown), indicating a pretty much identical performance of the two methods.

Consequently, our approach consistently outperforms the alternative of applying no smoothing when functions exhibit similarities over an interval, with low sensitivity to the choice of~$\alpha$, and leads to very similar performance when the functions are different. In the following simulation studies we set $\alpha=0.10$.

\subsection{Fixed designs}
\label{sec:SimFixedDesign}

\begin{figure}[t]
\centering
\includegraphics[width=0.48\textwidth]{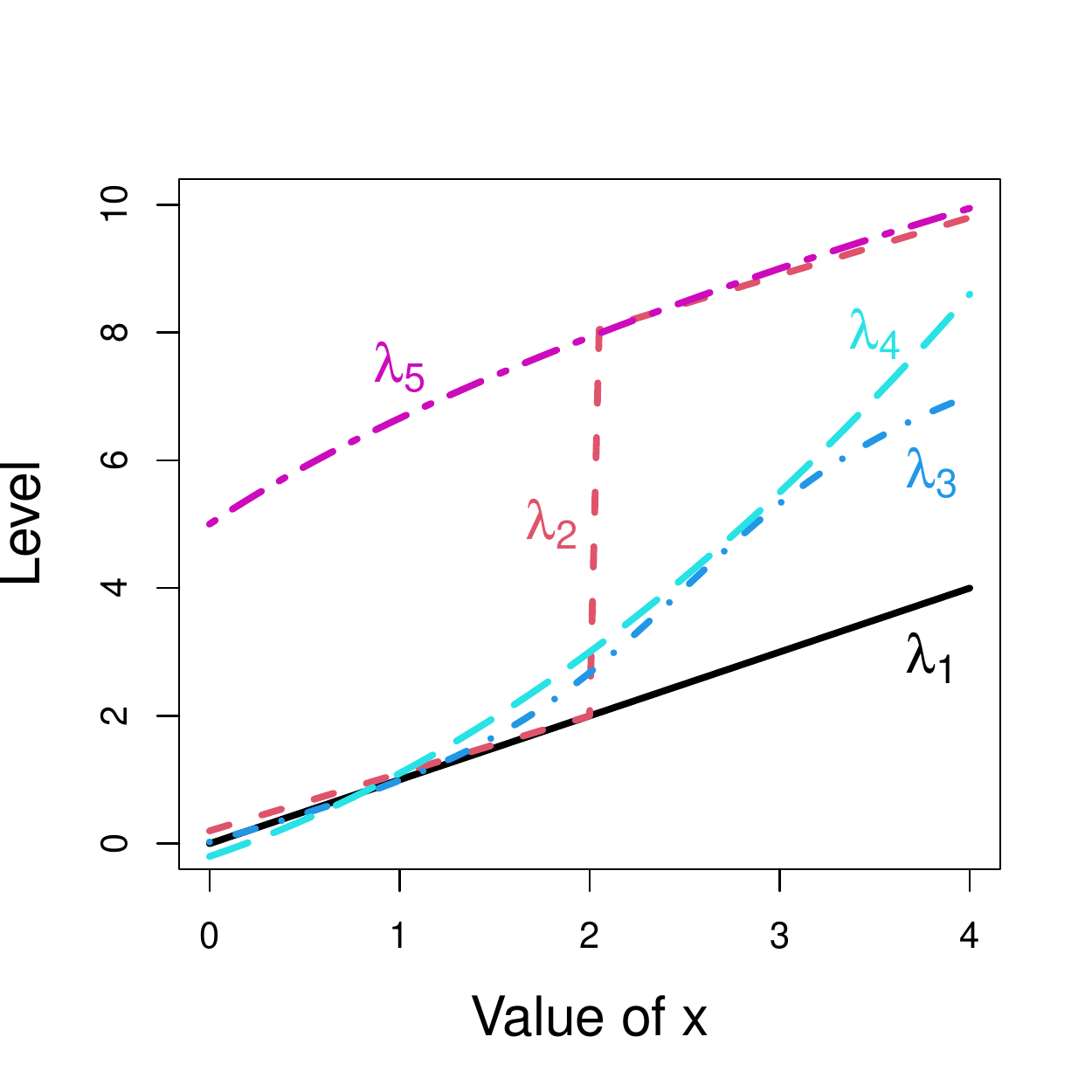}
\includegraphics[width=0.48\textwidth]{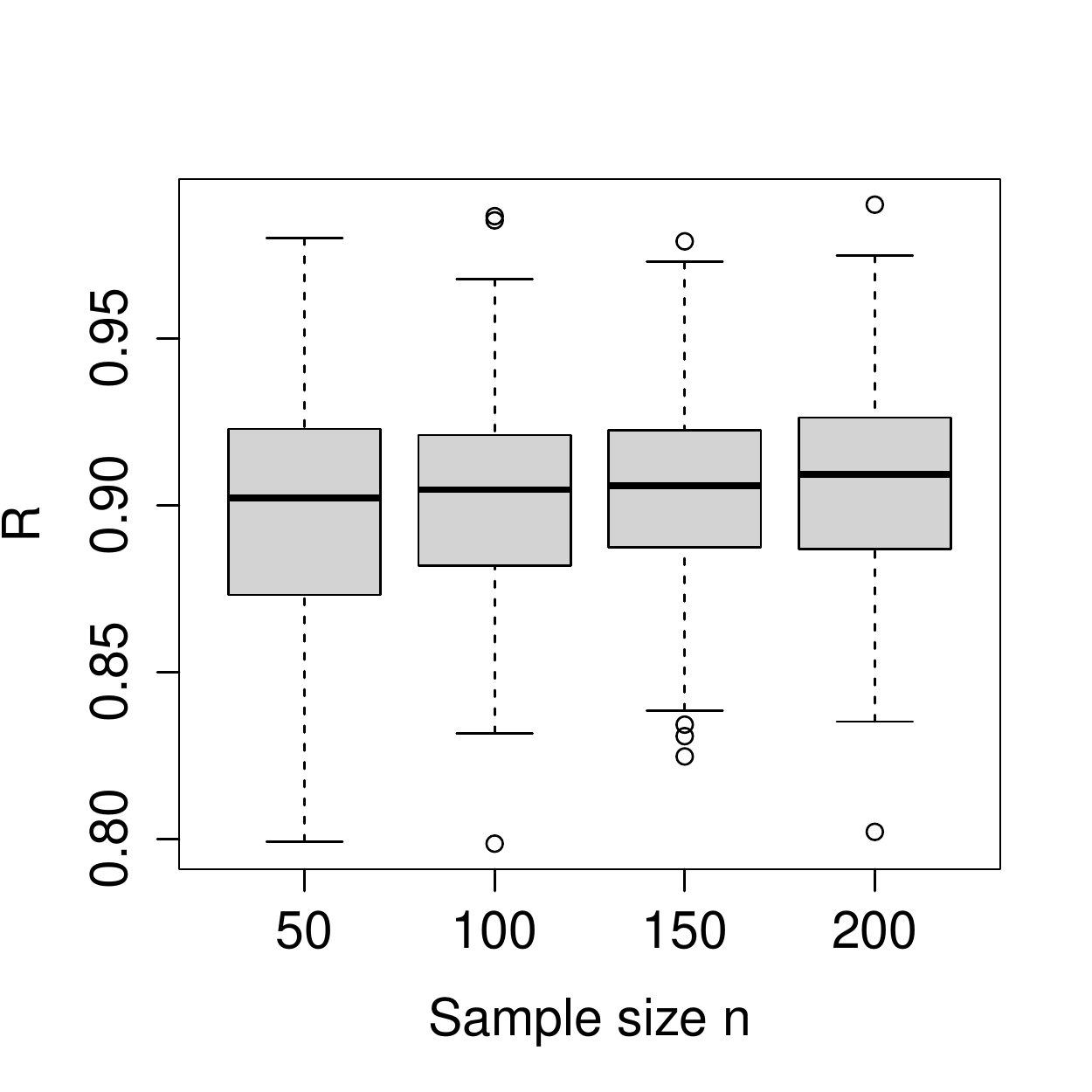}
\caption{Plot of the $K=5$ considered functions (left) and box plots of the performance measure $R$ for sample size $n\in\{50,100,150,200\}$ (right).}
\label{fig:FiveFunctions}
\end{figure}

After considering the case $m=1$ and $K=2$ we turn to more complex settings. Our first study considers $K=5$ functions over the interval $[0,4]$: $\lambda_1(x)=x$, $\lambda_2(x)=0.9x+0.2+6\mathbbm{1}(x>2)$, $\lambda_3(x)=0.8x+4\mathrm{expit}(2x-5)$, $\lambda_4(x)=0.3x^2+x-0.2$ and $\lambda_5(x) = 4\sqrt{x+1}+1$. Figure~\ref{fig:FiveFunctions} left panel illustrates that there is potential for information borrowing since each function is similar to at least one other function across a subset of $[0,4]$. As in Section~\ref{sec:Sensitivity} we consider sample sizes $n\in\{50,100,150,200\}$ and 200 data sets are generated for each value of $n$, with $x_i=4i/n$~($i=1,\ldots,n$), and $\epsilon_1,\ldots,\epsilon_5\sim\mathrm{Normal}(0,1)$ in \eqref{eq:GeneralIsoReg}.

\begin{table}[t]
\centering
\begin{tabular}{c|cc|cc|cc|cc|cc|cc}
\hline
$n$ &  \multicolumn{2}{|c|}{$\lambda_1$} &  \multicolumn{2}{|c|}{$\lambda_2$}&  \multicolumn{2}{|c|}{$\lambda_3$}&  \multicolumn{2}{|c|}{$\lambda_4$}&  \multicolumn{2}{|c|}{$\lambda_5$}&
\multicolumn{2}{|c}{Overall}\\
& $R$ & $P$ & $R$ & $P$ & $R$ & $P$ & $R$ & $P$ & $R$ & $P$ & $R$ & $P$\\
\hline
50 & 0.91&0.89 & 0.83&0.94 & 0.90&0.88 & 0.91&0.75 & 0.95&0.87 & 0.90&1.00 \\
100 & 0.90&0.88 & 0.85&0.98 & 0.87&0.87 & 0.93&0.81 & 0.94&0.91 & 0.90&1.00 \\
150 & 0.90&0.91 & 0.85&0.97 & 0.89&0.95 & 0.93&0.80 & 0.95&0.93 & 0.91&1.00 \\
200 & 0.89&0.87 & 0.86&0.97 & 0.89&0.92 & 0.94&0.82 & 0.94&0.93 & 0.91&1.00 \\
\hline
\end{tabular}
\caption{Median of $R$ and the estimate $P$ for the five functions in Figure~\ref{fig:FiveFunctions} left panel with sample size $n\in\{50,100,150,200\}$ and $\alpha=0.1$. The final two columns give the values for $R$ and $P$ as defined in Section~\ref{sec:Estimation}, while the remaining columns refer to performance for the individual functions.}
\label{tab:FiveFunctions}
\end{table}

Table~\ref{tab:FiveFunctions} shows that our approach consistently improves overall performance with $P=1$ for all considered sample sizes, that is, our approach improves the overall $R$ for all generated data sets. Figure~\ref{fig:FiveFunctions} right panel further shows that the improvement in performance is quite stable in terms of the interquartile range of $R$ for $n\in\{100,150,200\}$. When considering the results for the individual functions, we find marked differences. For example, the estimate for $\lambda_2$ improves in most cases, while a better estimate for $\lambda_4$ is only observed in 80\% of samples. Such a behaviour may be caused by the challenging configuration for higher values of $x$ - the estimate for $\lambda_4$ may be influenced by $\lambda_3$ or by $\lambda_2$ and $\lambda_5$. Further, the number of data points used in the test in Section~\ref{sec:SimilarityMeasure} decreases on average with the gradient of the function. Consequently, the weights for $\lambda_4$ and $x>3$ are more susceptible to randomness in the observed responses. This can also be seen by the values for $P$, which increase slowly with increasing $n$: we have more data available to correctly identify the differences between $\lambda_4$ and the remaining functions for $x>3$, while still using the data effectively to improve estimates for smaller values of $x$. 

We next turn to examples with $m=2$ explanatory variables. Two studies with $K=2$ functions are considered: in Study 1 $\lambda_1(\mathbf{x})=0.5x_1+0.7x_2 + 1.5 \mathbbm{1}\{\max_{m=1,2} x_m>2\}$ and $\lambda_2(\mathbf{x})=0.6x_1+0.5x_2 + 1.4\mathbbm{1}\{\max_{m=1,2} x_m>2\}$ are discontinuous at the same values of $\mathbf{x}$ and have generally not too different values, while in Study 2 $\lambda_1(\mathbf{x}) = 0.4 + 3\sqrt{(x_1/2)^2 + (x_2/2)^2-1}\,\mathbbm{1}\{(x_1/2)^2 + (x_2/2)^2 > 1\} $ is continuous and $\lambda_2(\mathbf{x}) = 0.3 +  \left\{0.3 + 4\sqrt{(2x_1/5)^2 + (2x_2/5)^2-1}\right\}\,\mathbbm{1}\{(2x_1/5)^2 + (2x_2/5)^2 > 1\}$ is discontinuous; a plot of these functions is provided in Section C in the online supplementary materials. Observations are sampled across a regular grid of varying size on $[0,4]^2$, and 200 data sets with $\epsilon_1,\epsilon_2\sim\mathrm{Normal}(0,1)$ are generated for each grid size. Table~\ref{tab:TwoDimensions} illustrates that our approach again improves overall performance for all generated data sets; the results for a third slightly modified version of Study 1 are provided in Section C of the online supplementary materials.

\begin{table}
\centering
\begin{tabular}{cc|cc|cc|cc}
\hline
Study & Grid size &  \multicolumn{2}{|c|}{$\lambda_1$} &  \multicolumn{2}{|c|}{$\lambda_2$}& 
\multicolumn{2}{|c}{Overall}\\
&& $R$ & $P$ & $R$ & $P$ & $R$ & $P$\\
\hline
&$10\times10$ &0.905 &0.965 &0.923 &0.925 &0.914 &1.000\\
1&$12\times12$ &0.899 &0.980 &0.933 &0.920 &0.914 &1.000\\
&$15\times15$ &0.910 &0.990 &0.938 &0.975 &0.922 &1.000\\
&$20\times20$ &0.922 &0.975 &0.939 &0.945 &0.930 &1.000\\
\hline
&$10\times10$ & 0.937 &0.985 &0.924 &0.995 &0.928 & 1.000\\
2&$12\times12$ & 0.942 &0.980 &0.921 &0.995 &0.930 & 1.000\\
&$15\times15$ & 0.942 &0.970 &0.920 &1.000 &0.932 & 1.000\\
&$20\times20$ & 0.954 &0.985 &0.925 &1.000 &0.939 & 1.000\\
\hline
\end{tabular}
\caption{Median of $R$ and the estimate $P$ for two pairs of functions for varying grid sizes and $\alpha=0.1$. The regular grid is spanned across the space $[0,4]\times[0,4]$.}
\label{tab:TwoDimensions}
\end{table}

Finally, we also applied our approach to binomially distributed response data with $K=2$ and $m=1$ and again find that our approach improves overall estimates; details are provided in Section D of the online supplementary materials.

In summary, the simulations demonstrate that our approach effectively borrows information across the data sets $\mathcal{D}_1,\ldots,\mathcal{D}_K$ in fixed designs and improves estimates when similarities between functions exist. The magnitude of improvement reduces with increasing sample size in certain settings, such as Study 2 in Table~\ref{tab:Sensitivity}, Section~\ref{sec:Sensitivity}. This may in part be caused by the slowly decreasing quantiles of the test statistic with increasing sample size $n$ under the null hypothesis - the chi-squared distribution with one degree of freedom likely provides a better approximation to the true distribution of the test statistic for small $n$, see Table~\ref{tab:DistirbutionLR} in Section~\ref{sec:SimilarityMeasure}.

\subsection{Random designs}
\label{sec:SimRandom}

After considering studies with fixed designs, Section~\ref{sec:SimFixedDesign}, we explore performance of the approach for random designs described in Section~\ref{sec:RandomDesign}. To aid comparability with the results for fixed designs, the $K=5$ functions $\lambda_1,\ldots,\lambda_5$ in the left panel of Figure~\ref{fig:FiveFunctions} are again considered. Observations are sampled independently with $Y_k\mid (X_k=x) \sim \mathrm{Normal}\{\lambda_k(x),1\}$ and $X_k\sim\mathrm{Uniform}(0,4)$~$(k=1,\ldots,5)$. Two random designs are studied. In the first design an equal number of observations is generated for $\mathcal{D}_1,\ldots,\mathcal{D}_5$, while in the second design $\mathcal{D}_2$ and $\mathcal{D}_3$ contain twice as many data points as $\mathcal{D}_1$, $\mathcal{D}_4$ and $\mathcal{D}_5$. The considered sample size $n$ for $\mathcal{D}_2$ (and $\mathcal{D}_3$) is $n_2\in\{50, 100, 150, 200\}$. We again calculate the performance measures $R$ and $P$ defined in Section~\ref{sec:Estimation}, but we only evaluate model fit for $\lambda_k$~($k=1,\ldots,5$) at the points in $\mathcal{D}_k$ to reduce sensitivity to the interpolation selected.

\begin{table}
\centering
\begin{tabular}{cc|cc|cc|cc|cc|cc|cc}
\hline
Design& $n_2$ &\multicolumn{2}{c}{$\lambda_1$} &\multicolumn{2}{|c|}{$\lambda_2$}&\multicolumn{2}{|c|}{$\lambda_3$}&\multicolumn{2}{|c|}{$\lambda_4$}&\multicolumn{2}{|c|}{$\lambda_5$}&
\multicolumn{2}{c}{Overall}\\
&& $R$ & $P$ & $R$ & $P$ & $R$ & $P$ & $R$ & $P$ & $R$ & $P$ & $R$ & $P$\\
\hline
&50 & 0.95&0.73 & 0.89&0.93 & 0.91&0.85 & 0.92&0.80 & 0.96&0.83 & 0.92&1\\
1&100 & 0.94&0.82 & 0.88&0.92 & 0.91&0.89 & 0.94&0.81 & 0.96&0.89 & 0.92&1\\
&150 & 0.92&0.87 & 0.89&0.97 & 0.92&0.90 & 0.95&0.74 & 0.96&0.89 & 0.93&1\\
&200 & 0.92&0.87 & 0.88&0.93 & 0.92&0.88 & 0.95&0.77 & 0.95&0.93 & 0.93&1\\
\hline
&50 & 0.92&0.74 & 0.92&0.91 & 0.94&0.85 & 0.87&0.92 & 0.94&0.86 & 0.92&1\\
2&100 & 0.90&0.81 & 0.92&0.91 & 0.94&0.85 & 0.89&0.87 & 0.92&0.89 & 0.91&1\\
&150 & 0.90&0.84 & 0.92&0.90 & 0.94&0.89 & 0.90&0.88 & 0.92&0.94 & 0.91&1\\
&200 & 0.89&0.91 & 0.93&0.94 & 0.94&0.89 & 0.92&0.85 & 0.91&0.94 & 0.92&1\\
\hline
\end{tabular}
\caption{Median of $R$ and the estimate $P$ for the five considered functions and the overall performance for the two considered random designs and $\alpha=0.1$. The column $n_2$ gives the number of data points in the set $\mathcal{D}_2$.}
\label{tab:FiveFunctionsRandom}
\end{table}

Table~\ref{tab:FiveFunctionsRandom} shows that our approach for information borrowing again consistently improves overall estimation performance. When comparing the results for the first design, Design 1, to the findings in Table~\ref{tab:FiveFunctions}, the improvement in terms of $R$ was slightly higher for the fixed than for the random design. As explained in Section~\ref{sec:RandomDesign}, our method may give $v_{k,p,i}=1$ in random designs, despite the function estimates only being similar. This potentially leads to oversmoothing at certain design points and explains the slightly worse performance in random designs. A comparison between the two designs shows that information is effectively borrowed across functions in both cases. In Design 2 the improvement in model fit is greater for the functions for which less data are available ($\lambda_1$, $\lambda_4$ and $\lambda_5$), but a consistent improvement is also achieved for $\lambda_2$ and $\lambda_3$. This difference is intuitive since information borrowing is expected to have increased capacity to improve estimation in settings with small sample size. Consequently, our approach effectively borrows information across data sets in a wide variety of designs.

\section{Case Studies}
\label{sec:CaseStudy}

\subsection{Neonatal mortality in Porto Alegre}

We apply our approach to investigate the relationship between infant birth weight and neonatal mortality (death within 7 days); comparing spatially structured neonatal risk groups. The data consist of the 183 neonatal deaths in Porto Alegre, Brazil, in 1998, together with 600 population based controls. Infants birth weight is recorded in grams and the data include georeferencing based upon the mother's residence at the time of the birth using upon the Universal Transverse Mercator (UTM). 

Previous analyses of the data \citep{Costain2009} indicate a reduced odds of neonatal mortality in the North-Westerly region of Porto Alegre after adjusting for infants birth weight. It is conceivable that the `residual' latent spatial structure is a reflection of spatially structured covariate effects not allowed for in the model, and moreover, that the effect of birth weight differs geographically, for example, due to socio-economic status and local infrastructure.

We split the data in to $K=2$ groups, referred to as lower risk ($k=1$) and higher risk ($k=2$), and model the form of the relationship between birth weight and neonatal mortality, allowing for non-linearity and threshold effects whilst borrowing information as appropriate. The number of observations in the lower and higher risk group is $n_1=114$ and $n_2=669$ respectively. Results from a t-test indicate that the two groups differ in their average birthweight, with an empirical difference of 200 grams.  

Let $y_{k,i}=1$ if the $i$-th infant in group $k$ survived 7 days and $y_{k,i}=0$ otherwise~($i=1,\ldots,n_k$). We model the response, conditional on the defined explanatory variable $X_k=$~birth weight, as $Y_k\mid(X_k=x)\sim\mbox{Bernoulli}\{\lambda_k(x)\}$, where $\lambda_k:\mathbb{N}\to[0,1]$ is a monotonically increasing function. Function estimates are derived using the approach described in Section~\ref{sec:RandomDesign} with $\alpha=0.1$.

\begin{figure}
\centering
\includegraphics[width=0.5\textwidth]{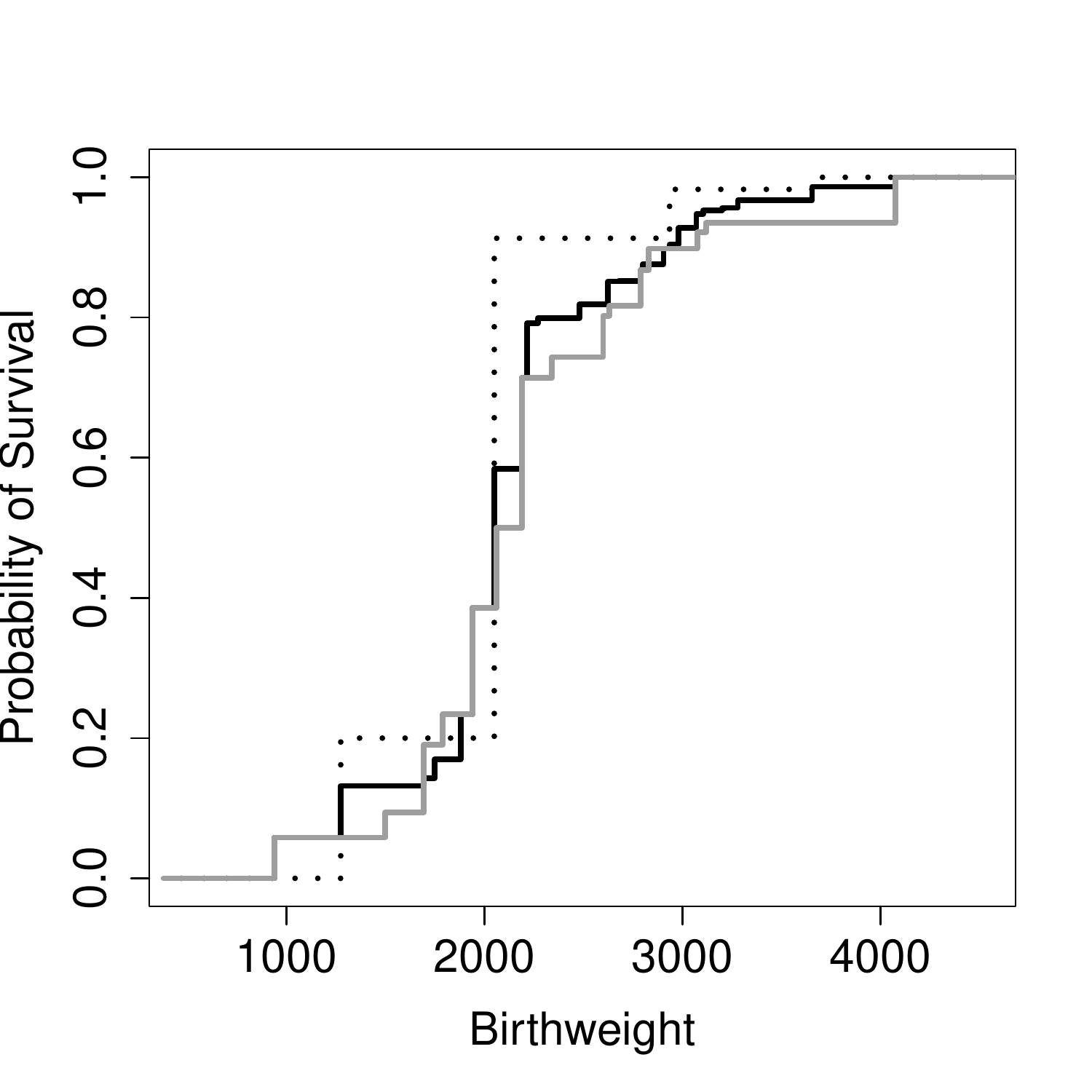}
\caption{Functions estimates for the lower risk group (black) and higher risk (grey) groups. The dotted line corresponds to the estimate obtained for the lower risk group using the approach by \cite{BarlowBrunk1972}.}
\label{fig:InfantsMortality}
\end{figure}

A number of insights can be drawn from the results in Figure~\ref{fig:InfantsMortality}. Infants with a birth weight below $1000~\mbox{grams}$ are unlikely to survive the neonatal period and there is a threshold effect at 2000 grams, the latter is not as clearly visible when a generalized additive model \citep{Wood2017} is fitted; see Section E in the online supplementary materials. Looking further at the functions, the estimate for $\lambda_1$ obtained using \cite{BarlowBrunk1972} is not very detailed, in particular for lower birth weights, since we only have 12 observations with birth weight below 2000 grams. By borrowing information from the high risk group, we obtain a more informed picture of the effect of birth weight on the risk of infant mortality, while still preserving the observed differences for higher birth weights.

\subsection{Stroke patient data from North West England}
\label{sec:stroke}

We analyse data on 87 female and 89 male stroke patients aged 60 and above from North West England who had a systolic blood pressure (SBP) of 170 mm Hg or above at time of admission; this is a subset of a data set with 484 patients aged 60 or above. On average, female patients were five years older than male patients and their SBP was similar. Our variable of interest is whether a patient was alive or deceased 28 days after admission, and we model the probability of death conditional on age and SBP, with a potentially varying regression function for female and male patients. Consequently, we have a setting with $K=2$ groups, female and male, and $m=2$ explanatory variables, age and SBP.

In the analysis we stratify patients according to age and SBP. The considered age brackets are 60--69, 70--79 and 80+, while observations for SBP are classified as 170--179~mm~Hg, 180--189~mm~Hg, 190--199~mm~Hg or $\geq$200~mm~Hg. This gives 12 possible combinations and we have at least 1 observation for each combination for male and female patients, that is, we have a fixed design with a varying number of observations across design points. To apply our approach, described in Section~\ref{sec:Method}, with $\alpha=0.1$, we assume that the probability of death is non-decreasing with increasing age and SBP. An exploratory analysis of the full data set with 484 patients suggests that the best survival rates are found for a SBP between 170 and 180 mm Hg and that probability of death increases beyond this level. 

\begin{table}
\centering
\begin{tabular}{ll|ccc|ccc}
\hline
Age & Blood Pressure & \multicolumn{3}{c|}{Women} & \multicolumn{3}{c}{Men}\\
&& $m$ & $\alpha=0.1$ & BB(1972) &$m$ & $\alpha=0.1$ & BB(1972)\\
\hline
      & 170--179  & 7 & 0.00 & 0.00 &8 &0.12 &0.13 \\
60-69 & 180--189  & 3 & 0.08 & 0.00 &12 &0.12 & 0.13\\
      & 190--199  & 1 & 0.09 & 0.00 &4 &0.12 & 0.13\\    
      & $\geq$200~~~     & 4 & 0.25 & 0.25 &11 &0.12 & 0.13\\    
\hline
      & 170--179  & 6  & 0.24 & 0.31 &10 &0.14 & 0.13\\
70-79 & 180--189  & 7  & 0.38 & 0.37 &7 &0.37 & 0.38\\   
      & 190--199  & 7  & 0.38 & 0.37 &9 &0.37 & 0.38\\   
      & $\geq$200~~~     & 12 & 0.38 & 0.37 &15 &0.40 & 0.44\\  
\hline      
      & 170--179 & 7  & 0.29 & 0.31 &3 &0.30 & 0.33\\
80+   & 180--189 & 9  & 0.38 & 0.37 &4 &0.40 & 0.44\\  
      & 190--199 & 8  & 0.38 & 0.38 &1 &0.40 & 0.44\\   
      & $\geq$200~~~    & 16 & 0.44 & 0.44 &5 &0.40 & 0.44\\
      \hline
\end{tabular}
\caption{Estimated probabilities for a patient not being alive 28 days after a stroke given their sex, age and blood pressure. The considered methods are the approach described in Section \ref{sec:Method} with $\alpha=0.1$ and the isotonic regression problem, BB(1972), by \cite{BarlowBrunk1972}. We also provide the number $m$ of data points per combination.}
\label{tab:Stroke}
\end{table}

Table~\ref{tab:Stroke} shows that the estimated probabilities of death are quite similar for female and male patients and not too different to the estimates derived using the isotonic regression problem. The largest differences can be found when the number of observations across the two groups vary the most in the 60-69 and 80+ categories. We can also see that probability of death noticeably increases in almost cases when comparing patients with 170-179 mm Hg to patients with SBP$\geq 180$, in particular at higher age. In summary, age and SBP appear to be factors contributing to whether a patient survives a stroke, while a patient's sex may affect the age at which they are most likely to suffer a stroke. 

\section{Discussion}
\label{sec:Discussion}

We considered the joint estimation of multiple monotonic functions when little, to no, information on their similarity is available. Our framework extended the isotonic regression problem by \citet{BarlowBrunk1972} and included a penalty for pairwise differences in the estimated function levels, with the penalty being tuned using a likelihood ratio test. Our numerical examples and case studies illustrated that our methodology leads to improved estimates if similarities between functions exist, and that it is robust otherwise. 

Our methodology may be used in applications different to that considered in Section~\ref{sec:CaseStudy}. As highlighted in the introduction, the estimation of dose-response across patient groups may be another option. Beyond these public health and pharmaceutics settings, our inference framework may be considered generally for estimating risk across several spatial sites, such as the risk of a person suffering from a heat stroke at various locations across Europe, with maximum daily temperature and length of the hot weather in days as explanatory variables. 

One small limitation of our framework is that we use the chi-squared distribution with one degree of freedom to derive the weights in~\eqref{eq:weight}. Since the rejection rate of the likelihood ratio test under equality of the functions is smaller than the specified significance level $\alpha$, specifying $\alpha$ is less straightforward due to a lack of a clear interpretation. A second limitation is that estimates tend to overfit for large $m$ - this is a common problem across isotonic regression due to a decreasing number of monotonicity constraints in high dimensions. Further, the power of our test in Section~\ref{sec:SimilarityMeasure} will likely decrease with increasing $m$. Some of these issues may be addressed by assuming an additive model structure as in \cite{Bergersen2014}, but the imposition of the constraint $\hat{y}_{k,t}=\hat{y}_{p,t}$ under the null hypothesis becomes more challenging

There are various ways to extend our methodology. Firstly one may investigate whether it is necessary to run the likelihood ratio test in Section~\ref{sec:SimilarityMeasure} for each data point in order to derive the weights \eqref{eq:weight}. While the computational cost was small in our case studies, it increases non-linearly with increasing $K$. We found that the computational cost tends to be higher when the functions are quite different, and thus we may want to avoid testing all points if there are strong indications of the functions being sufficiently different. A second extension may focus on replacing the considered loss function. While the quadratic loss function was used in \eqref{eq:OptimExtended}, due to computational reasons, the approach can also be applied with any other twice-differentiable strictly convex loss functions. Finally, one may consider setting an upper bound on $v_{k,p,i}~(k,p=1,\ldots,K;~i=1,\ldots,n)$. We set $v_{k,p,i}\leq 1$ but this may lead to oversmoothing for $K$ large. We could, for instance, impose $v_{k,p,i}\leq 1/(K-1)$ in order to give at least half of the weight to the observed data when updating the estimates for $\hat{\mathbf{y}}_k$ using \eqref{eq:UpdateLambdak}. 

\bibliography{sample}

\begin{thebibliography}{}

\bibitem[A{\i}t-Sahalia and Duarte, 2003]{Ait2003}
A{\i}t-Sahalia, Y. and Duarte, J. (2003).
\newblock Nonparametric option pricing under shape restrictions.
\newblock {\em Journal of Econometrics}, 116(1-2):9--47.

\bibitem[Bacchetti, 1989]{Bacchetti1989}
Bacchetti, P. (1989).
\newblock Additive isotonic models.
\newblock {\em Journal of the American Statistical Association},
  84(405):289--294.

\bibitem[Barlow and Brunk, 1972]{BarlowBrunk1972}
Barlow, R.~E. and Brunk, H.~D. (1972).
\newblock The isotonic regression problem and its dual.
\newblock {\em Journal of the American Statistical Association}, 67:140--147.

\bibitem[Bergersen et~al., 2014]{Bergersen2014}
Bergersen, L.~C., Tharmaratnam, K., and Glad, I.~K. (2014).
\newblock Monotone splines lasso.
\newblock {\em Computational Statistics \& Data Analysis}, 77:336--351.

\bibitem[Bowman et~al., 1998]{Bowman1998}
Bowman, A.~W., Jones, M.~C., and Gijbels, I. (1998).
\newblock Testing monotonicity of regression.
\newblock {\em Journal of Computational and Graphical Statististics},
  7:489--500.

\bibitem[Cai and Wei, 2021]{Cai2021}
Cai, T.~T. and Wei, H. (2021).
\newblock Transfer learning for nonparametric classification: Minimax rate and
  adaptive classifier.
\newblock {\em The Annals of Statistics}, 49(1):100--128.

\bibitem[Chatterjee et~al., 2018]{Chatterjee2018}
Chatterjee, S., Guntuboyina, A., and Sen, B. (2018).
\newblock On matrix estimation under monotonicity constraints.
\newblock {\em Bernoulli}, 24(2):1072--1100.

\bibitem[Costain, 2009]{Costain2009}
Costain, D.~A. (2009).
\newblock Bayesian partitioning for modeling and mapping spatial case–control
  data.
\newblock {\em Biometrics}, 65(4):1123--1132.

\bibitem[de~Leeuw et~al., 2009]{Leeuw2011}
de~Leeuw, J., Hornik, K., and Mair, P. (2009).
\newblock Isotone optimization in {R}: Pool-adjacent-violators algorithm
  {(PAVA)} and active set methods.
\newblock {\em Journal of Statistical Software}, 32(5):1--24.

\bibitem[Dunson, 2005]{Dunson2005}
Dunson, D. (2005).
\newblock Bayesian semiparametric isotonic regression for count data.
\newblock {\em Journal of the American Statistical Association},
  100(470):618--627.

\bibitem[{Durot}, 2002]{Durot2002}
{Durot}, C. (2002).
\newblock Sharp asymptotics for isotonic regression.
\newblock {\em Probability Theory and Related Fields}, 122(2):222--240.

\bibitem[Dykstra and Robertson, 1982]{Dykstra1982}
Dykstra, R.~L. and Robertson, T. (1982).
\newblock An algorithm for isotonic regression for two or more independent
  variables.
\newblock {\em The Annals of Statistics}, 10(3):708--716.

\bibitem[Fang et~al., 2021]{Fang2021}
Fang, B., Guntuboyina, A., and Sen, B. (2021).
\newblock Multivariate extensions of isotonic regression and total variation
  denoising via entire monotonicity and {Hardy--Krause} variation.
\newblock {\em The Annals of Statistics}, 49(2):769--792.

\bibitem[Gelfand and Kuo, 1991]{Gelfand1991}
Gelfand, A.~E. and Kuo, L. (1991).
\newblock Nonparametric {Bayesian} bioassay including ordered polytomous
  response.
\newblock {\em Biometrika}, 78:657--666.

\bibitem[Han et~al., 2019]{Han2019}
Han, Q., Wang, T., Chatterjee, S., and Samworth, R.~J. (2019).
\newblock Isotonic regression in general dimensions.
\newblock {\em The Annals of Statistics}, 47(5):2440--2471.

\bibitem[Hannah and Dunson, 2013]{Hannah2013}
Hannah, L.~A. and Dunson, D.~B. (2013).
\newblock Multivariate convex regression with adaptive partitioning.
\newblock {\em Journal of Machine Learning Research}, 14(1):3261--3294.

\bibitem[Hanson et~al., 1973]{Hanson1973}
Hanson, D.~L., Pledger, G., and Wright, F.~T. (1973).
\newblock On consistency in monotonic regression.
\newblock {\em The Annals of Statistics}, 1(3):401--421.

\bibitem[Lin and Dunson, 2014]{Lin2014}
Lin, L. and Dunson, D.~B. (2014).
\newblock {Bayesian monotone regression using {Gaussian} process projection}.
\newblock {\em Biometrika}, 101(2):303--317.

\bibitem[Luss and Rosset, 2014]{Luss2014}
Luss, R. and Rosset, S. (2014).
\newblock Generalized isotonic regression.
\newblock {\em Journal of Computational and Graphical Statistics},
  23(1):192--210.

\bibitem[Luss et~al., 2012]{Luss2012}
Luss, R., Rosset, S., and Shahar, M. (2012).
\newblock Efficient regularized isotonic regression with application to
  gene--gene interaction search.
\newblock {\em The Annals of Applied Statistics}, 6:253--283.

\bibitem[Meyer and Woodroofe, 2000]{Meyer2000}
Meyer, M.~C. and Woodroofe, M. (2000).
\newblock On the degrees of freedom in shape-restricted regression.
\newblock {\em The Annals of Statistics}, 28(4):1083--1104.

\bibitem[Nomakuchi and Shi, 1988]{Nomakuchi1988}
Nomakuchi, K. and Shi, N.-Z. (1988).
\newblock A test for a multiple isotonic regression problem.
\newblock {\em Biometrika}, 75(1):181--184.

\bibitem[Ramsay, 1998]{Ramsay1998}
Ramsay, J.~O. (1998).
\newblock Estimating smooth monotone functions.
\newblock {\em Journal of the Royal Statistical Society: Series B (Statistical
  Methodology)}, 60(2):365--375.

\bibitem[Reeve et~al., 2021]{Reeve2021}
Reeve, H.~W., Cannings, T.~I., and Samworth, R.~J. (2021).
\newblock Adaptive transfer learning.
\newblock {\em The Annals of Statistics}, 49(6):3618--3649.

\bibitem[Robertson and Wright, 1975]{Robertson1975}
Robertson, T. and Wright, F.~T. (1975).
\newblock Consistency in generalized isotonic regression.
\newblock {\em The Annals of Statistics}, 3:350--372.

\bibitem[Rohrbeck et~al., 2018]{Rohrbeck2017}
Rohrbeck, C., Costain, D.~A., and Frigessi, A. (2018).
\newblock Bayesian spatial monotonic multiple regression.
\newblock {\em Biometrika}, 105(3):691--707.

\bibitem[Royston, 2000]{Royston2000}
Royston, P. (2000).
\newblock A useful monotonic non-linear model with applications in medicine and
  epidemiology.
\newblock {\em Statistics in Medicine}, 19(15):2053--2066.

\bibitem[Saarela and Arjas, 2011]{Saarela2011}
Saarela, O. and Arjas, E. (2011).
\newblock A method for {B}ayesian monotonic multiple regression.
\newblock {\em Scandinavian Journal of Statistics}, 38(3):499--513.

\bibitem[Saarela et~al., 2023]{Saarela2020}
Saarela, O., Rohrbeck, C., and Arjas, E. (2023).
\newblock {Bayesian non-parametric ordinal regression Under a monotonicity
  constraint}.
\newblock {\em Bayesian Analysis}, 18(1):193 -- 221.

\bibitem[Sasabuchi et~al., 1983]{Sasabuchi1983}
Sasabuchi, S., M., I., and Kulatunga, D.~D.~S. (1983).
\newblock A multivariate version of isotonic regression.
\newblock {\em Biometrika}, 70(2):465--472.

\bibitem[Scott et~al., 2015]{Scott2015}
Scott, J.~G., Shively, T.~S., and Walker, S.~G. (2015).
\newblock Nonparametric {Bayesian} testing for monotonicity.
\newblock {\em Biometrika}, 102:617--630.

\bibitem[Shively et~al., 2009]{Shively2009}
Shively, T.~S., Sager, T.~W., and Walker, S.~G. (2009).
\newblock A {B}ayesian approach to non-parametric monotone function estimation.
\newblock {\em Journal of the Royal Statistical Society: Series B (Statistical
  Methodology)}, 71(1):159--175.

\bibitem[Tseng, 2001]{Tseng2001}
Tseng, P. (2001).
\newblock Convergence of a block coordinate descent method for
  nondifferentiable minimization.
\newblock {\em Journal of Optimization Theory and Applications},
  109(3):475--494.

\bibitem[Wilson et~al., 2014]{Wilson2014}
Wilson, A., Reif, D.~M., and Reich, B.~J. (2014).
\newblock Hierarchical dose-response modeling for high-throughput toxicity
  screening of environmental chemicals.
\newblock {\em Biometrics}, 70(1):237--246.

\bibitem[Wood, 2017]{Wood2017}
Wood, S. (2017).
\newblock {\em Generalized Additive Models: An Introduction with R}.
\newblock Chapman and Hall/CRC, Boca Raton, Fla., 2nd edition.

\bibitem[Wright, 1981]{Wright1981}
Wright, F.~T. (1981).
\newblock The asymptotic behavior of monotone regression estimates.
\newblock {\em The Annals of Statistics}, 9(2):443 -- 448.

\bibitem[Zhang, 2002]{Zhang2002}
Zhang, C.-H. (2002).
\newblock {Risk bounds in isotonic regression}.
\newblock {\em The Annals of Statistics}, 30(2):528 -- 555.

\end{thebibliography}

\newpage

\appendix
\renewcommand{\thesubsection}{\Alph{subsection}}

\section*{Online Supplementary Material}

\subsection{Power of the test}

\begin{figure}[h]
\centering
\includegraphics[width=0.49\textwidth]{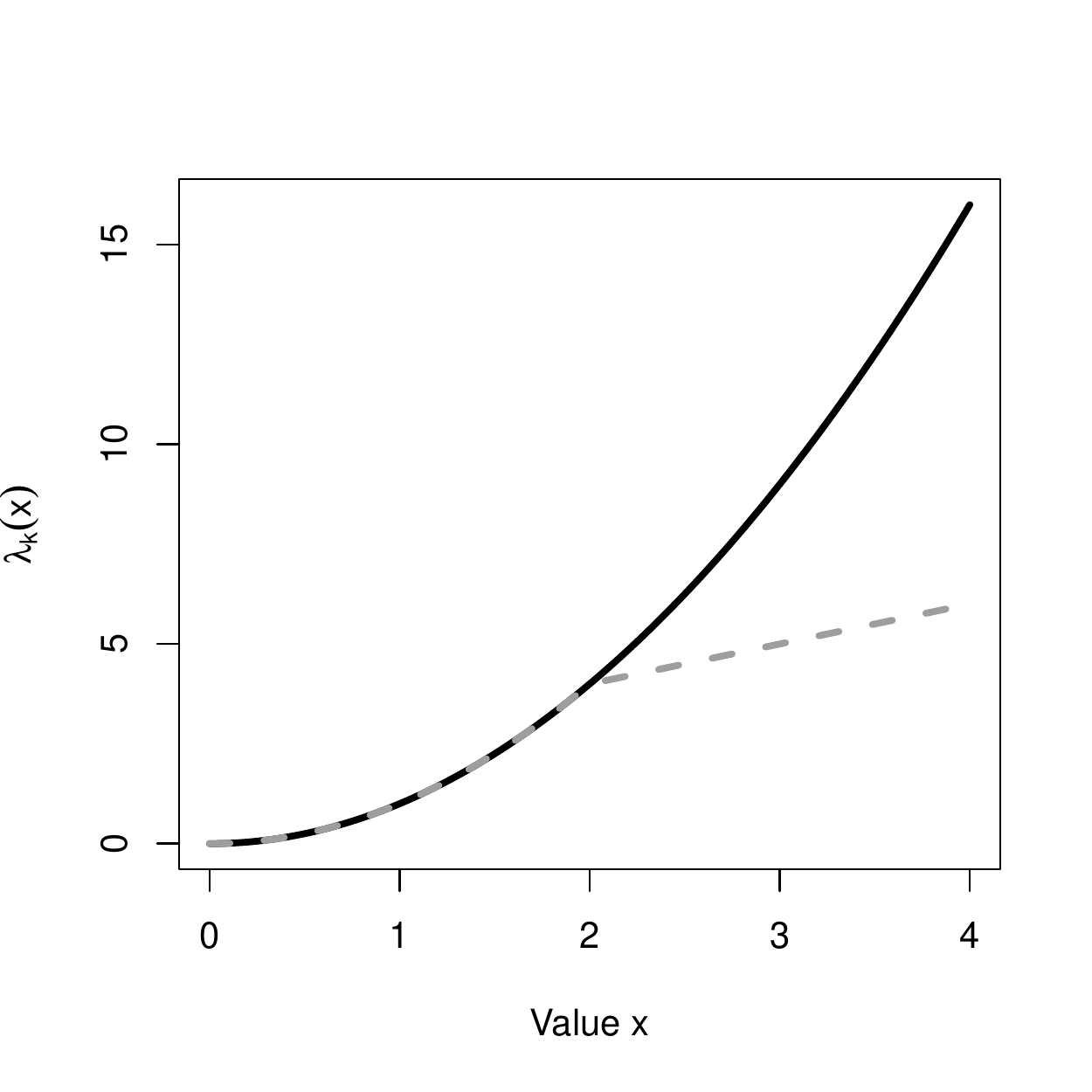}
\includegraphics[width=0.49\textwidth]{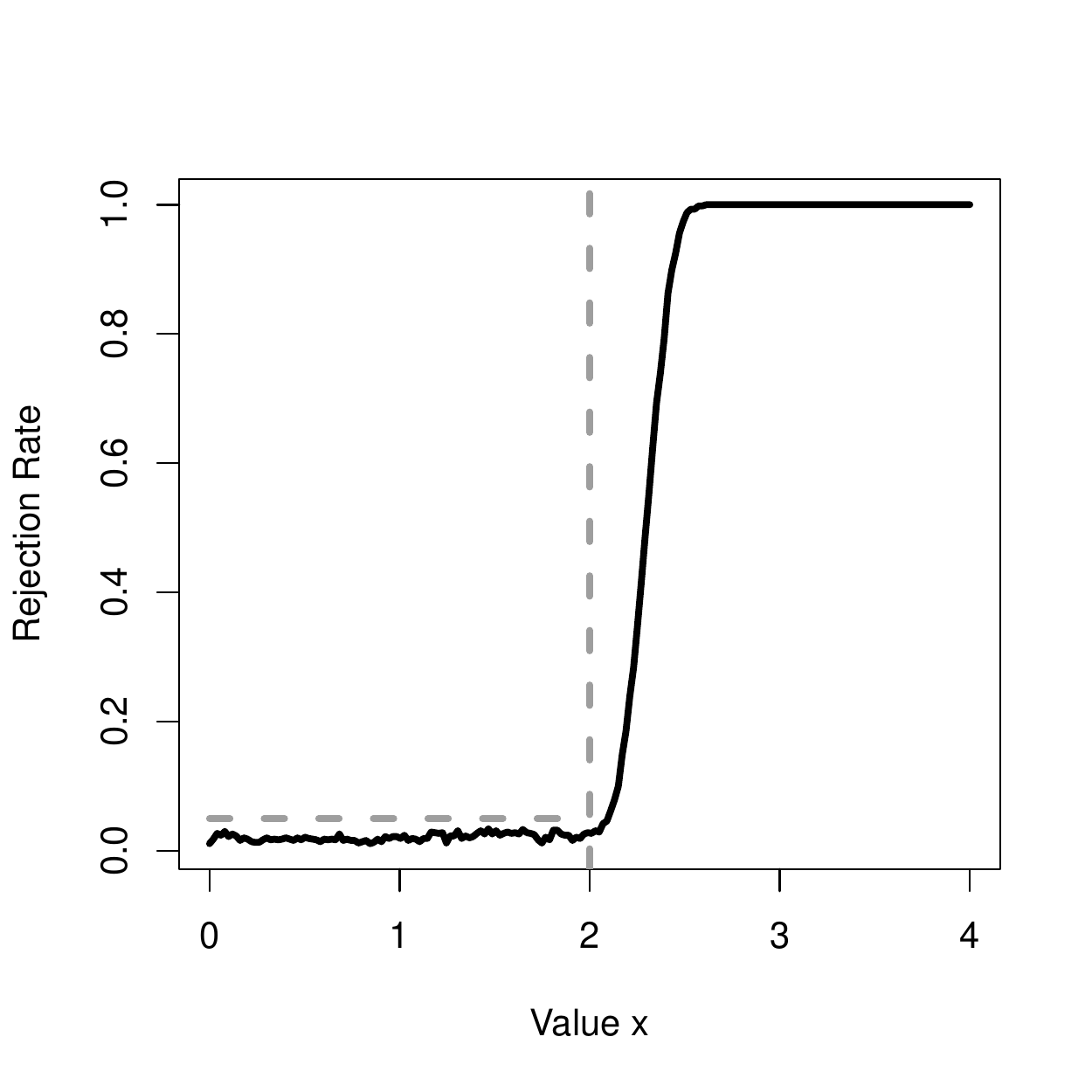}
\caption{Left panel: True functions $\lambda_1$ (black solid) and $\lambda_2$ (grey dashed). Right panel: Proportion of tests rejected at the values of $x$ (solid). The grey sahed lines represent the point beyond which the functions differ and a 5\% rejection rate.}
\label{fig:TestPower}
\end{figure}

We analyse performance of the proposed testing procedure using the functions $\lambda_1(x)=x^2$ and $\lambda_2(x) = x^2\mathbbm{1}\{x\leq2\} + (x+2) \mathbbm{1}\{x>2\}$ with $x\in[0,4]$ and $\epsilon_1,\epsilon_2\sim\mathrm{Normal}(0,1)$ in Equation (4) of the main paper. The left panel in Figure~\ref{fig:TestPower} shows that the functions are identical across the interval $[0,2]$ with differences in the levels increasing with increasing $x$ beyond this interval. We generated 1,000 data sets with $n=200$ equally spaced design points, $x_i=4i/n$~($i=1,\ldots,200$) and set $\alpha=0.05$ in the testing procedure, that is, we compare the test statistic LR$(k,p,t)$ to the 95\% quantile of a chi-squared distribution with one degree of freedom. 

The right panel in Figure~\ref{fig:TestPower} shows that the rejection rate is below the specified 5\% significance level for the points at which the functions are identical; this agrees with the empirical 95\% quantile of LR($k,p,t$) in Table 1 of the main paper lying below 3.84, the critical value we compare to. For values of $x>2$, the power of the test quickly increases with increasing difference between $\lambda_1(x)$ and $\lambda_2(x)$ and the null hypothesis is almost always rejected for $x>2.5$. This suggests that our test has a reasonable power despite being quite conservative.

\subsection{Distribution of the test statistic for random designs}

We again run a simulation study with $\lambda_k(x)=\lambda_p(x)=x^2$ and $\epsilon_k,\epsilon_p\sim\mathrm{Normal}(0,1)$. Observations for the explanatory variable are simulated independently from a Uniform(0,4) distribution and we calculate the test statistics LR$(k,p,t)$ at the empirical 25\%, 50\% and 75 \% quantiles of $X_1$. The sample size is $n\in\{50,200,1000\}$ and for each set up we generate 20,000 data sets.

\begin{table}[h]
\centering
\begin{tabular}{|lc|ccccc|}
\hline
Sample size& Quantile &$q_{0.5}$&$q_{0.7}$&$q_{0.8}$&$q_{0.9}$& $q_{0.95}$\\ 
\hline
         & 25\% &0.196 &0.550 &0.883 &1.554 &2.260\\
$n=50$   & 50\% &0.140 &0.490 &0.841 &1.503 &2.318\\
         & 75\% &0.088 &0.399 &0.728 &1.428 &2.214\\
\hline         
         & 25\% &0.247 &0.601 &0.924 &1.555 &2.239\\
$n=200$  & 50\% &0.224 &0.599 &0.947 &1.574 &2.262\\
         & 75\% &0.207 &0.561 &0.900 &1.550 &2.216\\
\hline         
         & 25\% &0.266 &0.622 &0.930 &1.540 &2.211\\
$n=1000$ & 50\% &0.255 &0.615 &0.945 &1.580 &2.218\\
         & 75\% &0.251 &0.617 &0.939 &1.569 &2.231\\
\hline         
\end{tabular}
\caption{Quantiles of $\mathrm{LR}(k,p,t)$ based on $20,000$ samples with $\lambda_k(x)=\lambda_p(x)=x^2$, sample size $n=\{50,200,1000\}$ and the empirical 25\%, 50\% and 75\% quantile of the explanatory variable.}
\label{tab:DistirbutionLRRandom}
\end{table}

Table~\ref{tab:DistirbutionLRRandom} indicates that the gradient has an affect on the distribution of $\mathrm{LR}(k,p,t)$, just as in Table~1 of the main paper, since the quantiles are slightly higher for the empirical 75\% quantile. The difference to the results for the fixed design is that the 95\% quantiles of LR$(k,p,t)$ are very similar across the considered samples sizes, and that the median is noticeably smaller for smaller sample sizes. The main reason for this difference is that there are quite a few cases with $\mathrm{LR}(k,p,t)=0$, in particular for small $n$. 

Table~\ref{tab:DistirbutionLRRandom2} shows the quantiles when conditioning on $\mathrm{LR}(k,p,t)>0$. We now again have that the 95\% quantile slowly decreases with increasing sample size $n$, just as for the fixed design in Table~1 of the main paper. Most importantly for our approach, the estimated quantiles again suggest that the upper quantiles of the distribution of $\mathrm{LR}(k,p,t)$ are smaller than that of a chi-squared distribution with one degree of freedom. Consequently, we again obtain a conservative test which we use to set the weight $v_{k,p,i}$ in Section 3.4 of the main paper. 

\begin{table}[ht]
\centering
\begin{tabular}{|lc|ccccc|}
\hline
Sample size& Quantile &$q_{0.5}$&$q_{0.7}$&$q_{0.8}$&$q_{0.9}$& $q_{0.95}$\\ 
\hline
         & 25\% &0.252 &0.618 &0.961 &1.641 &2.367\\ 
$n=50$   & 50\% &0.234 &0.623 &0.973 &1.684 &2.470\\ 
         & 75\% &0.197 &0.564 &0.913 &1.647 &2.478\\
\hline         
         & 25\% &0.265 &0.626 &0.955 &1.585 &2.272\\ 
$n=200$  & 50\% &0.256 &0.644 &0.993 &1.624 &2.317\\ 
         & 75\% &0.251 &0.621 &0.968 &1.622 &2.284\\ 
\hline         
         & 25\% &0.273 &0.629 &0.937 &1.545 &2.219\\ 
$n=1000$ & 50\% &0.265 &0.629 &0.961 &1.594 &2.232\\ 
         & 75\% &0.265 &0.636 &0.959 &1.593 &2.243\\ 
\hline         
\end{tabular}
\caption{Quantiles of $\mathrm{LR}(k,p,t)$ based on $20,000$ samples with $\lambda_k(x)=\lambda_p(x)=x^2$, sample size $n=\{50,200,1000\}$ and the empirical 25\%, 50\% and 75\% quantile of the explanatory variable. Compared to Table~\ref{tab:DistirbutionLRRandom}, we removed the cases where the added constraint had no effect and thus gave $\mathrm{LR}(k,p,t)=0$.}
\label{tab:DistirbutionLRRandom2}
\end{table}

\clearpage

\subsection{Examples with \texorpdfstring{$m=2$}{m} explanatory variables}

The following Figure~\ref{fig:SupplementTable4} shows the functions considered in the studies in Table~4 of the paper.

\begin{figure}[h]
\centering
\includegraphics[width=0.49\textwidth]{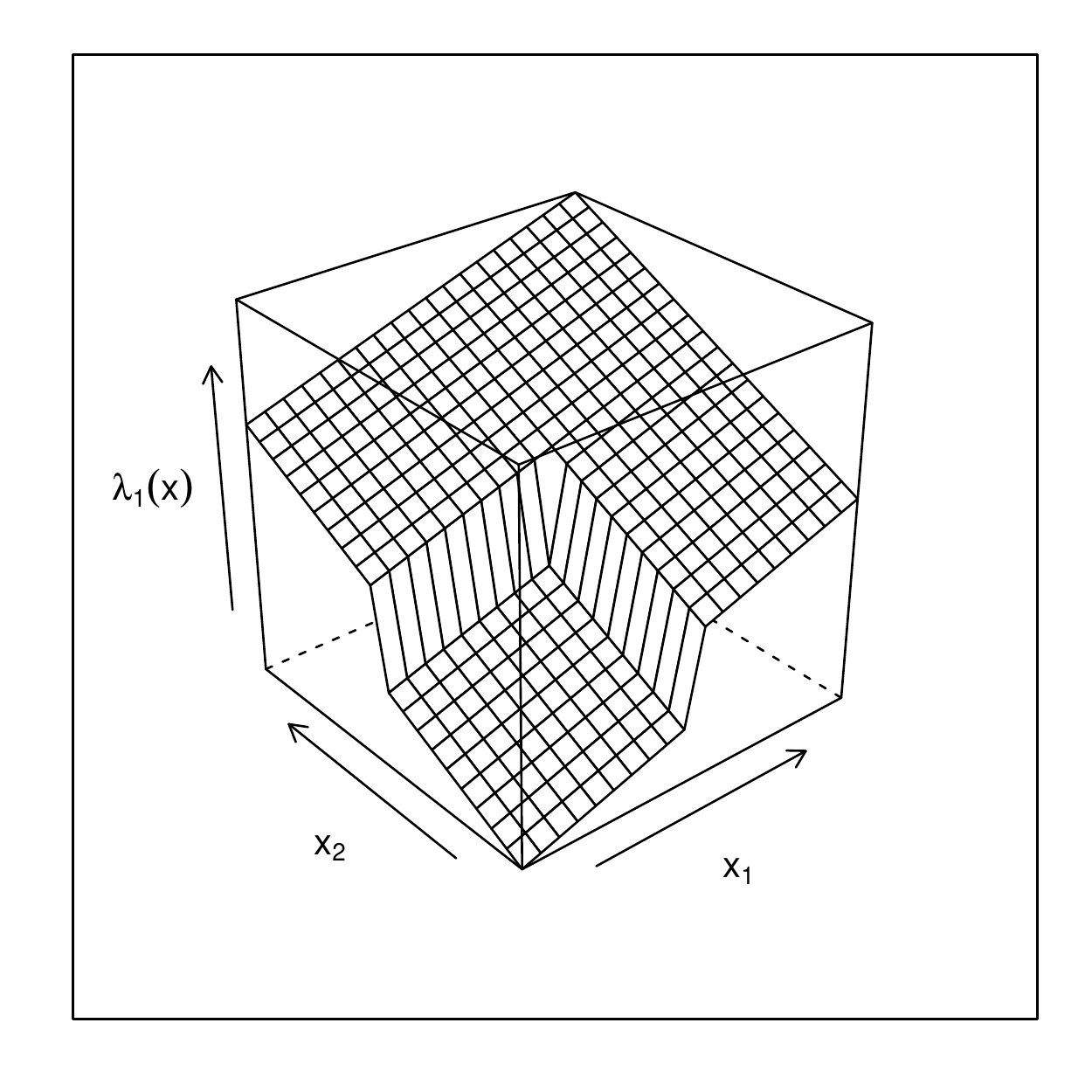}
\includegraphics[width=0.49\textwidth]{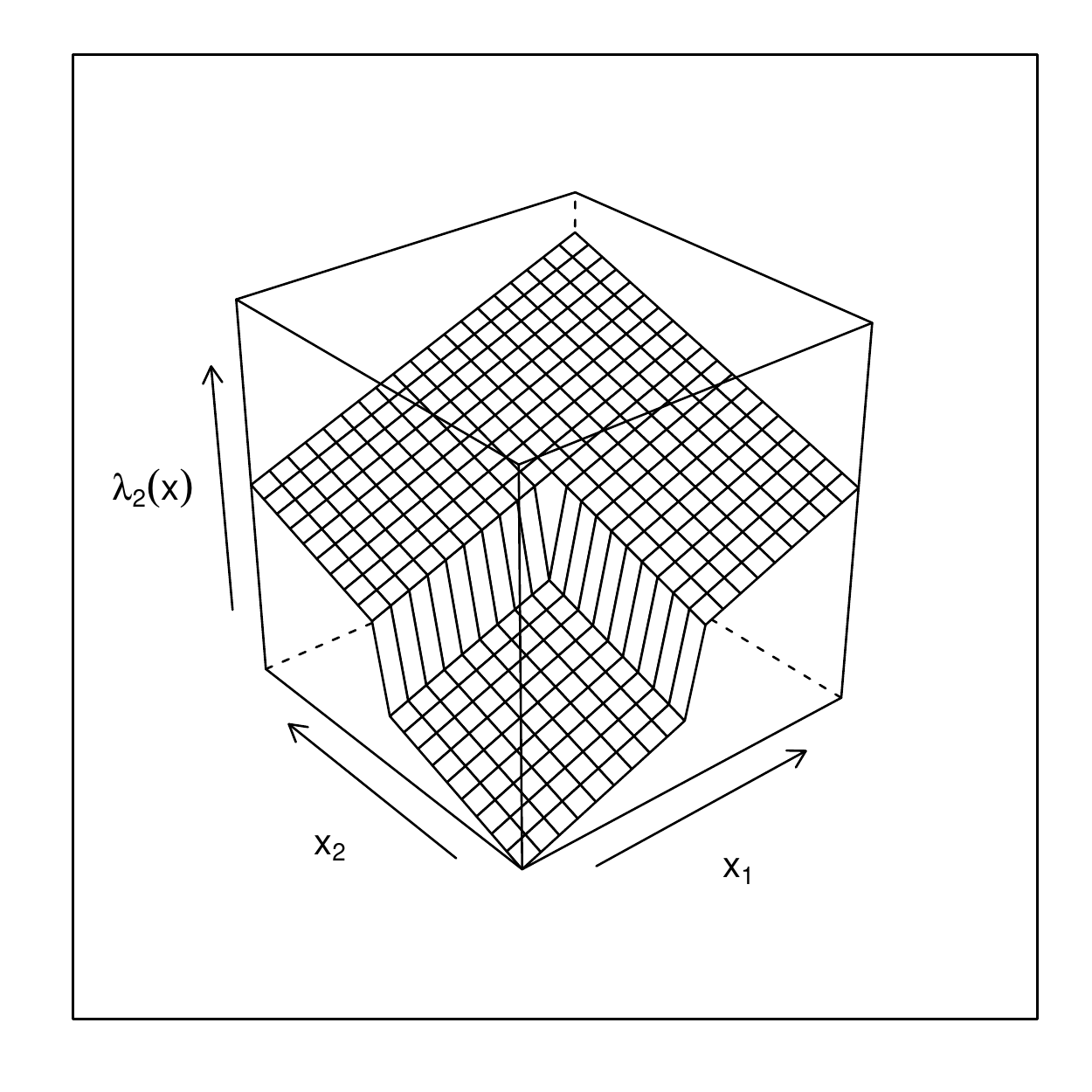}
\includegraphics[width=0.49\textwidth]{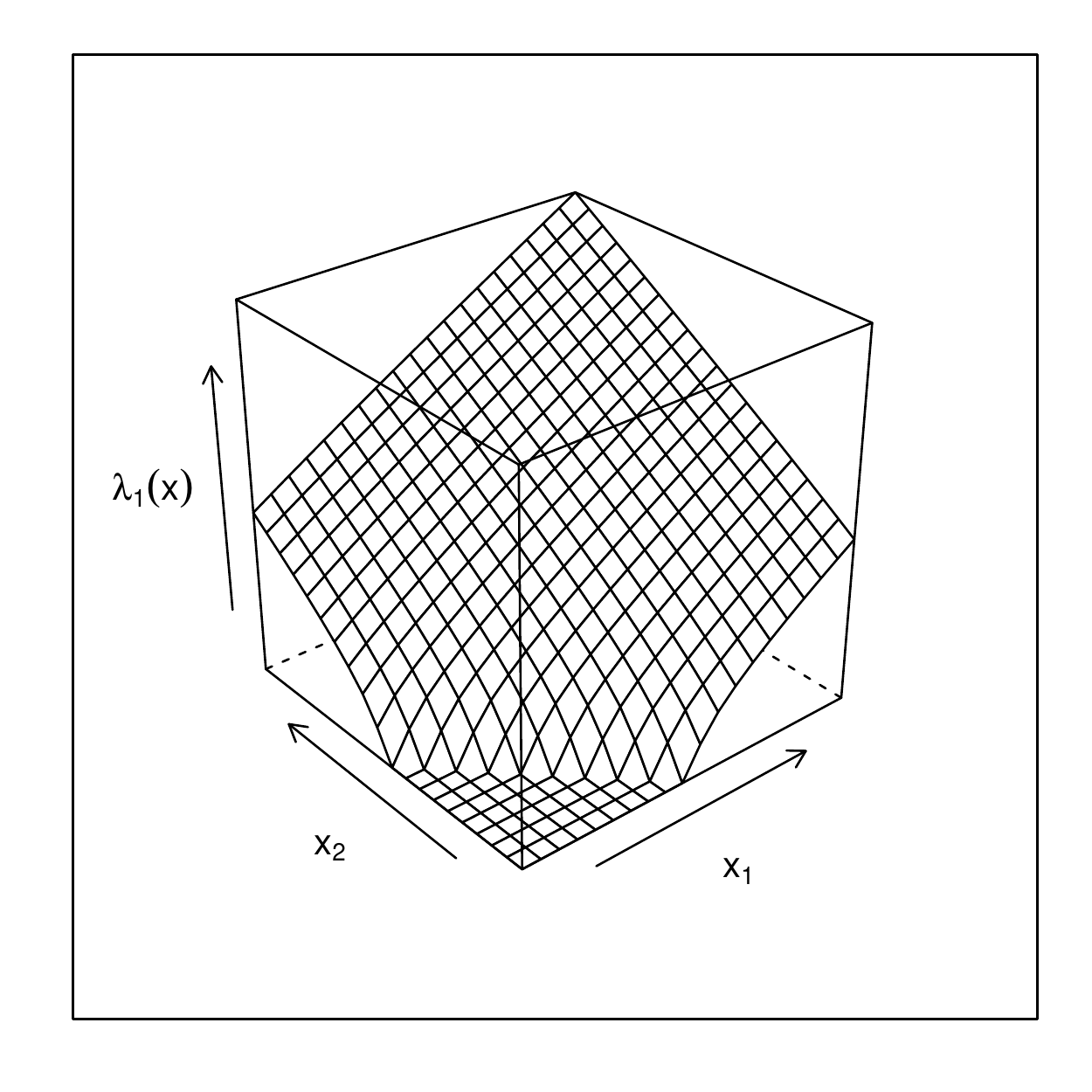}
\includegraphics[width=0.49\textwidth]{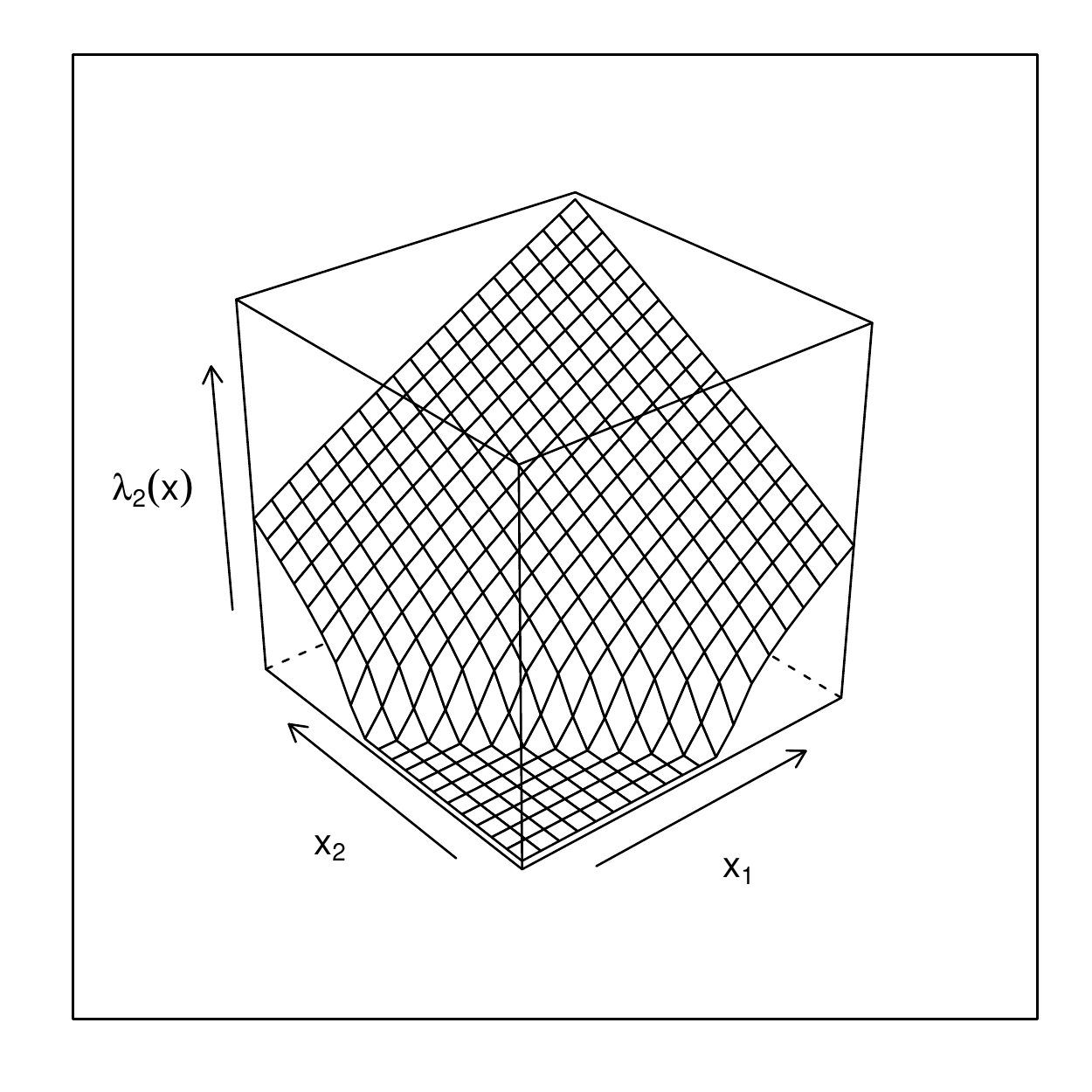}
\caption{Illustration of the function pairs $(\lambda_1,\lambda_2)$ considered in Table 4 in Section~4.2 of the main paper. The first row corresponds to Study 1, and the second row to Study 2, and all functions are plotted on a $20\times20$ regular grid.}
\label{fig:SupplementTable4}
\end{figure}

The following Table~\ref{tab:TwoDimensionsSupp} provides the results for a slightly modified setting to that in Study 1 of Table 4 in the main paper. While the two functions are less similar than that considered in Study 1, we again achieve a consistent improvement in the overall performance, in particular for smaller sample sizes.

\begin{table}
\centering
\begin{tabular}{|c|cc|cc|cc|}
\hline
Grid size &  \multicolumn{2}{|c|}{$\lambda_1$} &  \multicolumn{2}{|c|}{$\lambda_2$}& 
\multicolumn{2}{|c|}{Overall}\\
& $R$ & $P$ & $R$ & $P$ & $R$ & $P$\\
\hline
$10\times10$ &0.926 &0.930 &0.951 &0.835 &0.940 &0.995\\
$12\times12$ &0.927 &0.940 &0.959 &0.815 &0.941 &0.990\\
$15\times15$ &0.953 &0.875 &0.982 &0.710 &0.966 &0.940\\
$20\times20$ &0.955 &0.860 &0.980 &0.695 &0.968 &0.915\\
\hline
\end{tabular}
\caption{Median of $R$ and the estimate $P$ for
$\lambda_1(\mathbf{x})=0.5x_1+0.7x_2 + 1.5\mathbbm{1}\{\max_{m=1,2} x_m>2\}$ and
$\lambda_2=0.6x_1+0.5x_2 + 1.4 \mathbbm{1}\{\max_{m=1,2} x_m>1.5\}$ for varying grid sizes and $\alpha=0.1$.}
\label{tab:TwoDimensionsSupp}
\end{table}

\subsection{Numerical examples for binomial response data}

We start by considering an example with $K=2$ functions and $m=1$ explanatory variables. The functions are set to $\lambda_1(x) = 0.9\sin(x+0.3)+0.01$ and $\lambda_2(x) = \sin(x+0.2)$, $x\in(0,1)$. Observations for $Y_k$ are simulated from a binomial distribution with $Y_k\mid (X_k=x) \sim \mathrm{Binomial}\{5,\lambda_k(x)\}~(k=1,2)$. We consider the same set up as in Section~4.2 with 200 data sets generated for sample size $n\in\{50,100,150,200\}$ and estimates are derived with significance level $\alpha=0.1$. The set up is slightly different to that presented in Section 3.1 - we directly estimate the success probability $\lambda_k(x)$ and not the expectation $\mathbb{E}(Y_k\mid X=x)=5\lambda_k(x)$. While this makes no difference in terms of performance here, this aspect is important when considering set ups with varying sample sizes for which it is more natural to consider similarity in terms of the success probability, rather than the expectation.

\begin{table}[h]
\centering
\begin{tabular}{|c|cc|cc|cc|}
\hline
Sample size &\multicolumn{2}{|c|}{$\lambda_1$} &  \multicolumn{2}{|c|}{$\lambda_2$} & \multicolumn{2}{|c|}{Overall}\\ 
& $R$ & $P$ & $R$ & $P$ & $R$ & $P$\\
\hline
50  & 0.90\,(0.86,0.95) & 0.900 & 0.88\,(0.83,0.92) & 0.965 & 0.89\,(0.85,0.92) & 1.000\\
100 & 0.92\,(0.88,0.96) & 0.895 & 0.88\,(0.84,0.92) & 0.970 & 0.90\,(0.87,0.92) & 1.000\\
150 & 0.92\,(0.88,0.97) & 0.870 & 0.89\,(0.84,0.93) & 0.930 & 0.90\,(0.88,0.93) & 0.995\\
200 & 0.92\,(0.88,0.97) & 0.830 & 0.89\,(0.84,0.94) & 0.960 & 0.91\,(0.88,0.93) & 0.995\\
\hline
\end{tabular}
\caption{Performance measures $R$ (interquartile range) and $P$ for a setting with $K=2$ functions, binomially distributed response data and varying sample size $n\in\{50,100,150,200\}$.}
\end{table}

We next consider a setting with $K=4$ functions: $\lambda_1(x)=0.05+0.73\,\mathrm{expit}(10x-5)$, $\lambda_2(x)=0.2+0.55\,\mathrm{expit}(13x-4)$,
$\lambda_3(x)=0.2+0.7\,\mathrm{expit}(11x-4)$ and 
$\lambda_4(x)=0.05+0.8\,\mathrm{expit}(4x-4)$. If we interpret $\lambda_1,\ldots,\lambda_4$ as dose-response curves, we can say that the drug is quite effective for the groups corresponding to $\lambda_1$, $\lambda_2$ and $\lambda_3$, with varying optimal dose, while being less effective across the considered doses for the patient group with dose-response curve $\lambda_4$. We simulate data from $Y_k\mid (X_k=x) \sim \mathrm{Binomial}\{20,\lambda_k(x)\}~(k=1,\ldots,4)$, with the values of $x$ set as in the previous simulation study, that is, $x_i=i/n~(i=1,\ldots,n)$ and $n\in\{50,100,150,200\}$. In a second study we changed the distribution for the second group to $Y_2\mid (X_2=x) \sim \mathrm{Binomial}\{5,\lambda_2(x)\}$; we refer to this as the unbalanced design because it corresponds to an unequal number of samples being collected across groups. 

Table~\ref{tab:SimBinomK4} shows that we improve on the overall model fit for almost all simulated data sets, and that a higher rate of improvement is achieved in the unbalanced design, at the cost of the improvement being less consistent. Consequently, our approach is able to borrow information across groups to estimate functions for which less data are available, but there may be a risk of oversmoothing at small sample sizes.

\begin{table}[h]
\centering
\begin{tabular}{|c|cc|cc|}
\hline
& \multicolumn{2}{|c|}{Balanced Design} & \multicolumn{2}{|c|}{Unbalanced Design}\\
Sample size & $R$ & $P$ & $R$ & $P$ \\ 
\hline
50  & 0.96 (0.95,0.98) & 0.900 & 0.94 (0.90,0.99) & 0.805\\
100 & 0.96 (0.94,0.97) & 0.970 & 0.92 (0.87,0.95) & 0.920\\
150 & 0.95 (0.94,0.97) & 0.995 & 0.92 (0.88,0.95) & 0.960\\
200 & 0.96 (0.94,0.97) & 0.990 & 0.91 (0.87,0.94) & 0.965\\
\hline
\end{tabular}
\caption{Performance measures $R$ (interquartile range) and $P$ for a setting with $K=4$ functions, binomially distributed response data and varying sample size $n\in\{50,100,150,200\}$.}
\label{tab:SimBinomK4}
\end{table}

\subsection{Comparison to GAM for neonatal mortality data}

\begin{figure}[h]
\centering
\includegraphics[width=0.98\textwidth]{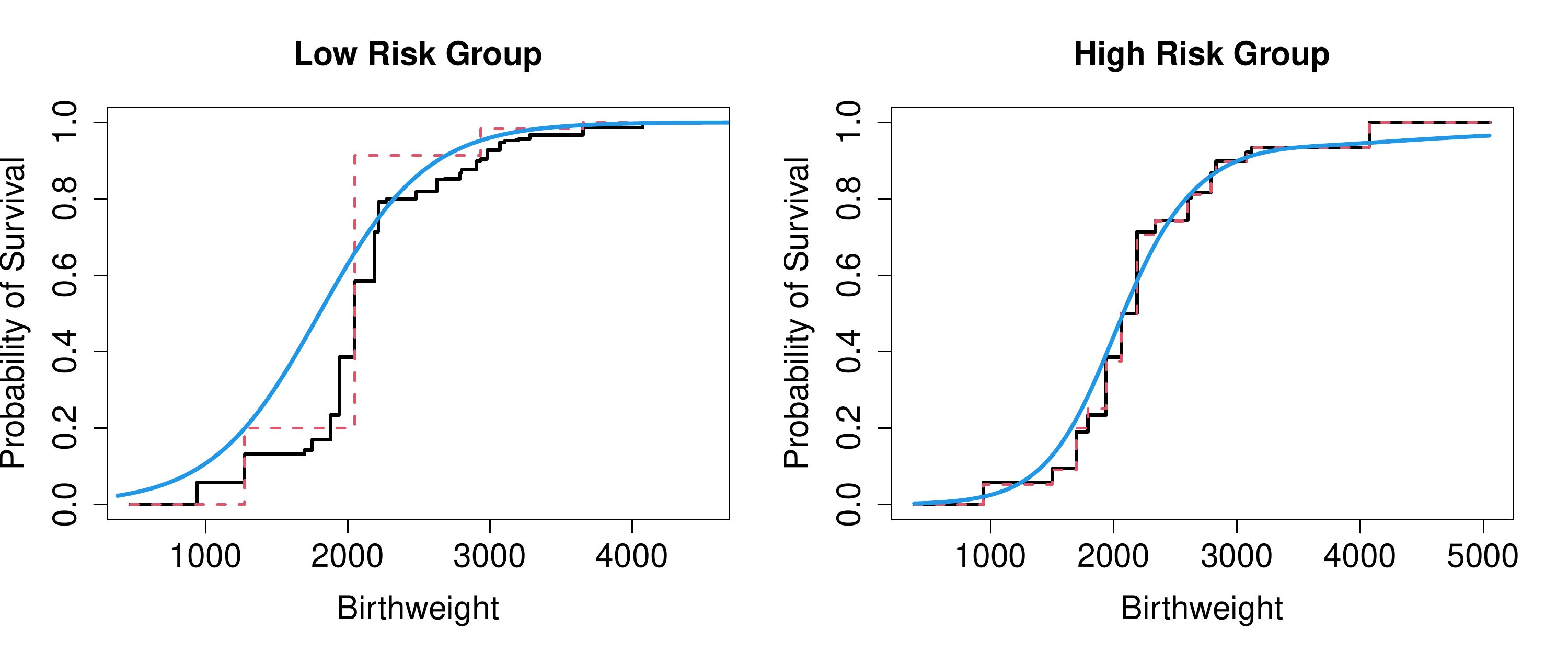}
\caption{Estimated regression curves for the low risk (left) and high risk (right) groups using the proposed approach (black), \citet{BarlowBrunk1972} (red) and a generalized additive model fitted (blue) which we fitted using the \texttt{mgcv} R package \citep{Wood2017}. The threshold effect close to 2000 grams is not clearly visible in the fitted generalized additive model.}
\label{fig:InfantComparison}
\end{figure}

\end{document}